# A fresh look at midpoint singularities in the algebra of string fields

**Theodore G. Erler**

*University of California, Santa Barbara*
*Santa Barbara, CA 93106, U.S.A*
*E-mail:*`terler@physics.ucsb.edu`

ABSTRACT: In this paper we study the midpoint structure of the algebra of open strings from the standpoint of the operator/Moyal formalism. We construct a split string description for the continuous Moyal product of hep-th/0202087, study the breakdown of associativity in the star algebra, and identify in infinite sequence of new (anti)commutative coordinates for the star product in in the complex plane. We also explain how poles in the open string non(anti)commutativity parameter correspond to certain "null" operators which annihilate the vertex, implying that states proportional to such operators tend to have vanishing star product with other string fields. The existence of such poles, we argue, presents an obstruction to realizing a well-defined formulation of the theory in terms of a Moyal product. We also comment on the interesting, but singular, representation $L_0$ which has appeared prominently in the recent studies of Bars *et al*.

# Contents

## 1. Introduction

Cubic string field theory began with a simple insight, due to Witten[1]: If a string field is imagined as a matrix whose indices correspond to the left and right halves of the string, matrix multiplication generates an associative algebra suitable for describing string interactions. This is a crucial insight, but Witten himself was the first to warn that it should be taken with a grain of salt. A string field is really a functional of a full and continuous string; to make it a matrix we would have to split the string into halves, which brings up all sorts of uncomfortable questions, for example how to set boundary conditions at the midpoint, how to treat the midpoint itself, what the left/right splitting does to the kinetic term, midpoint anomalies, etc. The ghost sector complicates the picture even more, since there star multiplication is not quite matrix multiplication even heuristically.

Yet, if Witten's product is *not* a local matrix product, what precisely *is it*? The answer remains somewhat unclear. The interaction vertex $\langle V_3|$, which is often taken as a working "definition" of the product, was explicitly constructed in ref.[2] using the matrix product intuition. However, once the precise form of $\langle V_3|$ had been determined, it seemed reasonable to leave the heuristic picture of "half string matrix multiplication" to the side, and work with concrete formulas. Still, it became clear with the discovery of associativity anomalies[3] and related problems in the context of the purely cubic string field theory[4] that, even with explicit formulas at hand, the singular nature of the Witten product could cause serious problems with any attempt to understand the theory analytically.

Recently, work in string field theory has been geared towards finding some evidence, either numerically or analytically, for Sen's well-known conjectures[5] regarding tachyon condensation. From the analytic side, the vacuum string field theory (VSFT) model[6, 7]—intended to describe physics around the locally stable closed string vacuum—has been particularly instructive. But because of the trivial form of the propagator, the interaction plays a defining role, and again questions regarding the singular nature of Witten's product are of central importance[6, 7, 8, 9, 10].

These issues are brought into particularly sharp relief by recent efforts which, despite mentioned difficulties, attempt a satisfactory definition of Witten's product explicitly as a half-string matrix or Moyal product. Many different approaches have been taken to this problem: some define the star product from the beginning as a matrix product—as is done in the split string[11, 12, 13] and discrete Moyal formalisms[14, 15, 16]—and others[17] attempt to "derive" such a representation from the $\langle V_3|$ itself. Although the starting points of the various approaches differ, it has become clear that they are all (almost) equivalent, except—crucially—their subtly differing treatment of the string midpoint.

In this paper, we attempt to bring all of these perspectives together and clarify the midpoint structure of Witten's product, particularly to understand precisely in what sense the algebra is—or is *not*—a matrix algebra. In section 1 we review the basic formulation of the star product as a Moyal product over a "phase space" of even coordinates and odd momenta, paying particular attention to domains and zero modes of the linear operator $T$ defining the "noncommutativity parameter" of Witten's product. In section 2 we consider the continuous or "kappa" basis[1] introduced in ref.[17], where the noncommutativity parameter is diagonal, and attempt to reformulate the (continuous) Moyal product as a half string matrix product. We find that the translation requires either the loss of a "string splitting" degree of freedom, or the introduction of an unphysical degree of freedom which allows the string to become disconnected from its midpoint. Therefore, we argue that Witten's product in the matter sector should properly be thought of as a Moyal product over phase space, rather than a matrix product over half strings. In section 3 we study ambiguities in the star product induced by its singular midpoint structure, as exemplified specifically by the Horowitz-Strominger "associativity anomaly."[3] We find that the nature of the anomaly depends in detail on how the star product is evaluated, and therefore should not be interpreted as a breakdown in associativity as such. Rather, the anomaly seems to indicate that we are multiplying string fields for which the star product is simply undefined—in particular, fields which depend on the unphysical "disconnected midpoint" degree of freedom. In section 4, we study the Moyal formulation of the ghost sector, which appears to be somewhat singular due to a pole in the non-anticommutativity parameter appearing in the kappa basis. We interpret this pole as giving rise to certain "null" operators which *annihilate* the vertex, implying in particular that states proportional to such operators tend to have vanishing star products with other string fields. Since an algebra which accommodates such states cannot have a well-defined identity, we interpret this as an obstruction to having a formulation of the theory in terms of a Moyal product. In section

---

[1]"kappa" here refers to a continuous parameter $\kappa \in [0, \infty]$ which labels the Moyal coordinates in the continuous basis.

5 we consider the possibility of Moyal coordinates labelled (in a distributional sense) by complex kappa. This allows us to discover an *infinite* sequence of new (anti)commutative coordinates for the open string star product, as well as an infinite sequence of null operators which annihilate the vertex. Finally, in section 6 we make some comments on an interesting, but singular representation for the Siegel gauge kinetic operator $L_0$ constructed by Bars *et al*[15]. We end with a few concluding remarks.

## 2. Basic concepts

A classical open string field $|\Psi\rangle$ is an element of the state space of a particular matter-ghost boundary conformal field theory(BCFT):

$$|\Psi\rangle \in \mathcal{H}_{\text{BCFT}} = \mathcal{H}_{matter} \otimes \mathcal{H}_{ghost}.$$

Typically, we are interested in the BCFT corresponding to an open string living on a space-filling D-25 brane, in which case we have Neumann boundary conditions in all directions at the string endpoints. Let us focus on the matter sector of the state space. We define "position" and "momentum" operators on the worldsheet in terms of the standard mode oscillators $\alpha_n$:

$$x(\sigma) = x_0 + \sqrt{2} \sum_{n=1}^{\infty} x_n \cos n\sigma \qquad x_n = \frac{i}{n\sqrt{2}}(\alpha_n - \alpha_{-n})$$

$$\pi p(\sigma) = p + \sqrt{2} \sum_{n=1}^{\infty} p_n \cos n\sigma \qquad p_n = \frac{1}{\sqrt{2}}(\alpha_n + \alpha_{-n}),$$

where $[\alpha_m, \alpha_n] = m\delta_{m,-n}$ (suppressing Lorentz indices). The modes $x_n$ and $p_n$ have eigenstates:

$$\langle x_n| = \langle 0|\mathcal{O}(x_n), \qquad \mathcal{O}(x_n) = \exp\left[-\frac{n}{2}(x_n^2 - 2\sqrt{2}i\alpha_n x_n - \alpha_n^2)\right]$$

$$\langle p_n| = \langle 0|\tilde{\mathcal{O}}(p_n), \qquad \tilde{\mathcal{O}}(p_n) = \exp\left[-\frac{1}{2n}(p_n^2 - 2\sqrt{2}\alpha_n p_n + \alpha_n^2)\right],$$

where $x_n, p_n$ now refers to the eigenvalue and $|0\rangle$ is the (matter sector) $SL(2,\mathbb{C})$ vacuum. In order to formulate the open string star product as a Moyal product, we will find it necessary to represent the string field as a functional of the *even* position Fourier modes and the *odd* momentum Fourier modes (restricting ourselves to zero momentum[2]):

$$\Psi[x_{2n}, p_{2n-1}] = \langle 0| \prod_{n=1}^{\infty} \mathcal{O}(x_{2n})\tilde{\mathcal{O}}(p_{2n-1})|\Psi\rangle. \qquad (2.1)$$

It is useful to think of the modes $x_{2n}, p_{2n-1}$ as defining a choice of coordinates (sometimes called the "mixed basis") on a linear space $\mathbb{P}$. A string field is then simply complex valued functional on $\mathbb{P}$. Vectors in $\mathbb{P}$ have a natural inner product:

$$\langle u, u' \rangle = \sum_{n=1}^{\infty} \left[\frac{1}{2n-1}p_{2n-1}p'_{2n-1} + 2n x_{2n} x'_{2n}\right], \qquad u, u' \in \mathbb{P}$$

---

[2]We make this simplification since most of the issues we will discuss are associated with large mode numbers

induced by considering string configurations on which fields in the perturbative Fock space have nonvanishing support[3].

The open string star product of functionals on $\mathbb{P}$ can be calculated with the formula,

$$\Psi * \Phi(u) = \mathcal{N} \, \Psi \star \Phi(u),$$

where $\mathcal{N}$ is a normalization[4] and $\star$ is a canonically normalized Moyal product satisfying,

$$[x_{2n}, p_{2m-1}]_\star = 2i \, T_{2n,2m-1}, \tag{2.2}$$

with the other Moyal commutators vanishing. The matrix $T$ above is given by ,

$$\begin{aligned} T_{2n,2m-1} &= \frac{4}{\pi} \int_0^{\pi/2} d\sigma \cos 2n\sigma \cos(2n-1)\sigma \\ &= \frac{2(-1)^{m+n+1}}{\pi} \left( \frac{1}{2m-1+2n} + \frac{1}{2m-1-2n} \right). \end{aligned} \tag{2.3}$$

The string modes $x_{2n}, p_{2n-1}$ are only a particular basis for $\mathbb{P}$. Clearly, it is worthwhile to search for another basis in which the matrix elements of $T$ take a simpler form. More on this in a moment.

We may think of $T$ as a linear operator between two (identical) Hilbert spaces: one of "odd moded" sequences, $\mathcal{H}_{odd}$, and the other of "even moded" sequences $\mathcal{H}_{even}$:

$$T : \mathcal{H}_{odd} \to \mathcal{H}_{even},$$

with,

$$\begin{aligned} \langle a, b \rangle &= \sum_{n=1}^\infty n a_{2n-1} b_{2n-1} \quad a, b \in \mathcal{H}_{odd} \\ \langle s, t \rangle &= \sum_{n=1}^\infty n s_{2n} t_{2n} \quad\quad\quad s, t \in \mathcal{H}_{even}. \end{aligned} \tag{2.4}$$

The Sobolev-type inner product on $\mathcal{H}_{odd}$ and $\mathcal{H}_{even}$ make these spaces *smaller* than the usual Hilbert space of square summable sequences, $\ell^2$. As an operator on $\mathcal{H}_{odd}$, $T$ has an inverse[14, 9], an operator often called $R$ in the literature:

$$\begin{aligned} R_{2m-1,2n} &= \frac{4}{\pi} \int_0^{\pi/2} d\sigma \cos(2m-1)\sigma [\cos 2n\sigma - (-1)^n] \\ &= \frac{4n(-1)^{n+m}}{\pi(2m-1)} \left( \frac{1}{2m-1+2n} - \frac{1}{2m-1-2n} \right) \\ &= \frac{(2n)^2}{(2n-1)^2} T_{2n,2m-1}, \end{aligned} \tag{2.5}$$

---

[3]Specifically, the vacuum state takes the form $\Psi_{|0\rangle}(u) = \exp(-\frac{1}{2}\langle u, u \rangle)$.

[4]This normalization constant is infinite, even in $D = 26$ when ghosts are taken into account[18]. It's not clear how this should be interpreted, or how much of a problem it represents. It was suggested in ref.[17] that the infinite normalization might be an intrinsic fault in the Witten vertex, and other vertices would be more satisfactory. The cost of choosing a different vertex for open string field theory is severe, however: the deformed algebra would be *nonassociative* (only homotopy associative—an $A_\infty$ algebra) and the action would be *non-polynomial*[19]. At this point there seems to be no reason to take such drastic measures to fix up this constant.

and
$$R : \mathcal{H}_{even} \to \mathcal{H}_{odd}, \qquad RT = 1_{odd}, \quad TR = 1_{even}. \tag{2.6}$$

On the other hand, we can use the matrix elements of $T$ to define, by extension, a linear operator $T$ (using the same name) on the Hilbert space of sequences $D(T)$ for which the sum $\sum_{n=1}^{\infty} T_{2m,2n-1} a_{2n-1}$ is finite. The linear space $D(T)$ is quite a bit larger than $\mathcal{H}_{odd}$. Most importantly, $T$ has a zero mode in $D(T)$ [9],

$$v_{2n-1} = \frac{2\sqrt{2}}{\pi} \frac{(-1)^{n+1}}{2n-1}, \quad Tv = 0, \tag{2.7}$$

which is not normalizable[5] in $\mathcal{H}_{odd}$, consistent with the fact that $T$ is invertible there. Apparently, on $D(T)$ the equation $RT = 1$ is no longer true. This phenomenon is easy to understand by analogy to the equation $\frac{1}{x}x = 1$. When acting on a suitably well-behaved space of functions, the equation makes perfect sense; but if it were to act on a delta function, for example, the equation is wrong: $\frac{1}{x}(x\delta(x)) = 0, \neq \delta(x)$. Given the fact that $RT = 1$ in $\mathcal{H}_{odd}$, we can by extension define $RT = 1$ on all of $D(T)$; but then, in a sense, we loose associativity in operator/vector multiplication,

$$R(Tv) \equiv RTv = 0 \quad (RT)v = v.$$

These "associativity anomalies"[9] are related to midpoint ambiguities in the star algebra, and play an important role in our analysis.

Alternatively, we define the transpose operator $\bar{T}$ which maps from even moded sequences to odd moded sequences. In this case it is natural to define the even/odd mode Hilbert spaces $\mathcal{H}'_{even}$ and $\mathcal{H}'_{odd}$ to be $-\frac{1}{2}$ order Sobolev spaces. On $\mathcal{H}'_{even}$, $\bar{T}$ has an inverse, $\bar{R}$; but outside of $\mathcal{H}'_{odd}$, $\bar{R}$ has a zero mode,

$$v'_{2n-1} = \frac{2\sqrt{2}}{\pi}(2n-1)(-1)^{n+1}, \tag{2.8}$$

as is readily seen from eq.(2.6). Again we run into a tricky zero mode.

Let us consider a few other choices of basis for which the star product takes a simpler form. Two sets of bases, sometimes called the "discrete even" and the "discrete odd" bases, have been used extensively in the work of Bars and collaborators[14, 15]:

$$\text{even}: \quad x_{2n} = x_{2n}^D \qquad\qquad p_{2n-1} = \tfrac{1}{2}\sum_{m=1}^{\infty} p_{2m}^D T_{2m,2n-1}$$

$$\text{odd}: \quad x_{2n-1}^D = \tfrac{1}{2}\sum_{m=1}^{\infty} R_{2n-1,2m} x_{2m} \qquad p_{2n-1}^D = p_{2m-1}. \tag{2.9}$$

Substituting these relations into eq.(2.2), bearing in mind eq.(2.6), we find[6]

$$[x_{2n}^D, p_{2m}^D]_\star = i\delta_{2n,2m} \qquad [x_{2n-1}^D, p_{2m-1}^D]_\star = i\delta_{2n-1,2m-1}.$$

---

[5] The normalization is chosen, as in [9] so that $\sum_{n=1}^{\infty} v_{2n-1}^2 = 1$.

[6] Most of the work in the discrete Moyal formalism has focused on the even basis, probably for historical reasons[14]. The odd basis is just as good.

An important assumption here is that the transformations eq.(2.9) are invertible. They are indeed invertible if the contracted and free indices in eq.(2.9) operate in the correct Hilbert spaces. The contracted index does not pose a problem, since $(x_{2n}, p_{2n-1})$ are elements of $\mathbb{P}$, which has the right inner product for eq.(2.9) to be invertible. The free index on the other hand is more illusive. In a string functional it will appear contracted somehow, and we know little about the Hilbert space on which it acts. Presumably, the Hilbert space of the free index would be determined by a suitable definition of the algebra of string fields, though it is not yet clear whether the definition is likely to allow the change of basis eq.(2.9). It was shown for instance in ref.[15] that the perturbative vacuum is a well-defined functional after the change of basis eq.(2.9) but the sliver state[20] is not.

Whatever the precise definition of the star algebra ends up being, it is clear that at least sometimes it is useful to think about fields for which the transformation eq.(2.9) is not invertible. This is where problems arise. Consider, for example, the half-string momentum functional rewritten in the even basis:

$$P_L = \int_0^{\pi/2} d\sigma p(\sigma) = \frac{\sqrt{2}}{\pi} \sum_{n=1}^{\infty} \frac{(-1)^{n+1}}{2n-1} p_{2n-1} \implies P_L = \sum_{n=1}^{\infty} 0 p_{2n}^D = 0. \qquad (2.10)$$

This vanishing of $P_L$ is a direct consequence of the fact that $T$ has a zero mode. In ref.[9] this pathology was explained as arising from the fact that the discrete even basis imposes Neumann boundary conditions at the midpoint, making it impossible to describe string configurations with a midpoint discontinuity. The discrete odd basis has an analogous problem:

$$D = \frac{2\sqrt{2}}{\pi} \sum_{n=1}^{\infty} \frac{(-1)^{n+1}}{2n-1} x_{2n-1}^D \implies D = \sum_{n=1}^{\infty} 0 x_{2n} = 0. \qquad (2.11)$$

This singularity occurs because the discrete odd basis, which imposes Dirichlet boundary conditions at the midpoint, has more degrees of freedom than can be ascribed to a "reasonable" string[9].

A strategy for dealing with these problems is to regulate/deform $T$ into an invertible operator so that the basis eq.(2.9) becomes well defined. This approach has been developed extensively by Bars *et al* in references [9, 14, 15, 16, 21, 22, 23]. This approach has advantages, as will be explored for instance in Section 6, but it is based on the assumption that $T$ in some sense "should be" invertible; our goal is to avoid such assumptions, so we will not follow their perspective too closely here. We do not believe that $T$ needs to be invertible *a priori* in order to give string field theory a satisfactory definition.

We should mention another basis on $\mathbb{P}$ which has sometimes been discussed[24, 16], the so-called "string bit" basis:

$$x^b(\sigma) = \sqrt{\frac{8}{\pi}} \sum_{n=1}^{\infty} x_{2n-1}^D \cos(2n-1)\sigma \quad p^b(\sigma) = \sqrt{\frac{8}{\pi}} \sum_{n=1}^{\infty} p_{2n-1} \cos(2n-1)\sigma, \qquad (2.12)$$

for $\sigma \in [0, \frac{\pi}{2}]$, satisfying,

$$[x^b(\sigma), p^b(\sigma')]_\star = 2i\delta(\sigma - \sigma').$$

This basis may prove useful for studying tensionless strings[24, 16], but is otherwise problematic because it does not unambiguously impose open string boundary conditions. Equation (2.12) defines the boundary conditions implicitly, but other mode expansions could be used and the basis would have identical appearance. To see why this is a problem, consider the tachyon mass-shell constraint:

$$\left(L_0\left[x(\sigma), \frac{\delta}{\delta x(\sigma)}\right] - 1\right)\langle x(\sigma)|p\rangle = 0.$$

This equation is not meaningful unless we consistently expand $x(\sigma)$ in terms of functions with a particular set of boundary conditions appropriate for the BCFT in question. Otherwise, for example, we could expand $x(\sigma)$ with Neumann boundary conditions in $L_0$, but then with periodic boundary conditions in $\langle x(\sigma)|p\rangle$; the above equation will then clearly not work. Therefore the bit basis can only work effectively when expanded in a set of functions with specific boundary conditions, such as in eq.(2.12). But of course, in doing this we are not really working in the bit basis anymore.

## 3. Split strings and the continuous basis

Having described the background, we turn to the continuous kappa basis for the Moyal product. This approach was originally derived from the zero momentum three string vertex $|V_3\rangle$ [2], specifically by choosing a basis of oscillators which diagonalized the quadratic form defining the squeezed state expression of the vertex[25]. In this form, the vertex factorized into a continuous tensor product of vertices, each of which described a Moyal product. This approach has since been extended to include zero modes[26, 27], ghosts[28, 18], superstring interactions[29], and interactions in the presence of NS-NS flux[30]. In the original form of [17] the kappa basis is defined by the coordinates,

$$x(\kappa) = \sqrt{2}\sum_{n=1}^{\infty} v_{2n}(\kappa)\sqrt{2n}\,x_{2n}$$

$$y(\kappa) = -\sqrt{2}\sum_{n=1}^{\infty} \frac{v_{2n-1}(\kappa)}{\sqrt{2n-1}}\,p_{2n-1}, \tag{3.1}$$

where $\kappa \in [0, \infty)$ and the functions $v_n(\kappa)$ are defined implicitly through the generating function[25],

$$\sum_{n=1}^{\infty}\frac{z^n}{\sqrt{n}}v_n(\kappa) = \frac{1 - e^{-\kappa \tan^{-1}z}}{\kappa N(\kappa)}, \quad N(\kappa) = \left(\frac{2}{\kappa}\sinh\frac{\pi\kappa}{4}\right)^{1/2}. \tag{3.2}$$

In ref.[31] the $v_n(\kappa)$'s were shown to be complete and orthonormal on the interval $\kappa \in (-\infty, \infty)$. In fact, noting that $v_n(-\kappa) = (-1)^{n+1}v_n(\kappa)$ we can write:

$$\sum_{m=1}^{\infty} v_{2m-1}(k)v_{2m-1}(k') = \tfrac{1}{2}\delta(k-k') \qquad \int_0^{\infty} dk\; v_{2m-1}(k)v_{2n-1}(k) = \tfrac{1}{2}\delta_{2m-1,2n-1}$$

$$\sum_{m=1}^{\infty} v_{2m}(k)v_{2m}(k') = \tfrac{1}{2}\delta(k-k') \qquad \int_0^{\infty} dk\; v_{2m}(k)v_{2n}(k) = \tfrac{1}{2}\delta_{2m,2n}. \tag{3.3}$$

These equations only strictly hold when $\kappa > 0$. More information about these special functions can be found in the appendix. Equation (3.1) is a unitary change of basis on $\mathbb{P}$, with the inner product in the two bases related by,

$$\sum_{n=1}^{\infty} \left[ \frac{1}{2n-1} p_{2n-1} p'_{2n-1} + 2n x_{2n} x'_{2n} \right] = \int_0^{\infty} d\kappa [x(\kappa) x'(\kappa) + y(\kappa) y'(\kappa)].$$

After invoking the formula[17],

$$T_{2m-1,2n} = -2\sqrt{\frac{2m-1}{2n}} \int_0^{\infty} d\kappa \ \tanh \tfrac{\pi \kappa}{4} v_{2m-1}(\kappa) v_{2n}(\kappa), \tag{3.4}$$

it is simple to show from eq.(2.2) that the Moyal coordinates satisfy commutation relations,

$$[x(\kappa), y(\kappa')]_\star = 2i \tanh \tfrac{\pi \kappa}{4} \delta(\kappa - \kappa'). \tag{3.5}$$

The noncommutativity parameter[7] $\theta(\kappa) = \tanh \frac{\pi \kappa}{4}$ is diagonal, but its magnitude varies as a function of $\kappa$. Most significantly, $\theta(\kappa)$ vanishes continuously toward $\kappa = 0$, so the associated coordinates become commutative. At $\kappa = 0$, $x(\kappa)$ vanishes while $y(\kappa)$ becomes proportional to the half-string momentum, eq.(2.10). The authors of [17] considered this a compelling argument in favor of the $\kappa$ basis, since in a naive treatment of the discrete basis the significance of this coordinate is easily missed.

We now attempt to gain further insight into the structure of the kappa basis by translating it into an (almost) equivalent split string description, where the role of the midpoint is more apparent. An extensive treatment of the half string formalism in the kappa basis appears in ref.[32].

Often it is convenient to represent multiplication in the algebra of functions on noncommutative $\Re^2$ as a nonlocal Moyal product between otherwise commutative functions on $\Re^2$. However, another useful perspective is the "split string" matrix-like representation of the algebra. This can be related to the Moyal perspective as follows: Given a function $\psi(x,p)$ with $[\hat{x}, \hat{p}] = 2i\theta$, define,

$$\hat{\psi}(\theta z + x, -\theta z + x) = \bar{\psi}(z + \tfrac{x}{\theta}, -z + \tfrac{x}{\theta}) = \int_{-\infty}^{\infty} dp e^{ipz} \psi(x, p).$$

A simple excercise[14] shows that calculating the "matrix product" of these objects,

$$\hat{A} \star \hat{B}(l,r) = \theta \int_{-\infty}^{\infty} dw \hat{A}(l,w) \hat{B}(w,r)$$

$$\bar{A} \star \bar{B}(l,r) = \int_{-\infty}^{\infty} dw \bar{A}(l,w) \bar{B}(w,r), \tag{3.6}$$

is equivalent to calculating the Moyal product of $A(x,p)$ and $B(x,p)$. Note that when $\theta \neq 1$ there are two "natural" split string representations of the product. We distinguish between these by writing fields alternatively with a hat or with a bar. The significance of this distinction will be clear in a moment.

---

[7]To avoid ugly factors of $\frac{1}{2}$ we will define the noncommutativity parameter so that $[x,p] = 2i\theta$, so that it differs by a factor of one half from the more standard convention.

We can easily follow this recipe to construct a split string representation of the kappa basis. Write,

$$\Psi[x_n] = \int \prod_{n=1}^{\infty} \left( \frac{dp_{2n-1}}{2\pi} e^{ip_{2n-1}x_{2n-1}} \right) \Psi[x_{2n}, p_{2n-1}]$$
$$= \int [dy(\kappa)]_1 \exp\left( \int_0^\infty d\kappa y(\kappa) z(\kappa) \right) \Psi^M[x(\kappa), y(\kappa)]$$
$$= \hat{\Psi}[l(\kappa), r(\kappa)] = \bar{\Psi}[\bar{l}(\kappa), \bar{r}(\kappa)],$$

with

$$l(\kappa) = \theta(\kappa) z(\kappa) + x(\kappa), \qquad r(\kappa) = -\theta(\kappa) z(\kappa) + x(\kappa)$$
$$\bar{l}(\kappa) = z(\kappa) + \frac{x(\kappa)}{\theta(\kappa)} \qquad \bar{r}(\kappa) = -z(\kappa) + \frac{x(\kappa)}{\theta(\kappa)}. \tag{3.7}$$

We can then calculate the open string star product as,

$$\hat{A} * \hat{B}[l(\kappa), r(\kappa)] = \mathcal{N} \int [dw(\kappa)]_2 \hat{A}[l(\kappa), w(\kappa)] \hat{B}[w(\kappa), r(\kappa)]$$
$$\bar{A} * \bar{B}[\bar{l}(\kappa), \bar{r}(\kappa)] = \mathcal{N} \int [dw(\kappa)]_3 \bar{A}[\bar{l}(\kappa), w(\kappa)] \bar{B}[w(\kappa), \bar{r}(\kappa)]. \tag{3.8}$$

The measures in the previous formulae are normalized so that,

$$1 = \int [dw(\kappa)]_1 \exp\left[ -\frac{1}{2\pi} \int_0^\infty d\kappa d\kappa' K_{-1}(\kappa, \kappa') w(\kappa) w(\kappa') \right]$$
$$= \int [dw(\kappa)]_2 \exp\left[ -\frac{\pi}{2} \int_0^\infty d\kappa d\kappa' K_{-1}(\kappa, \kappa') w(\kappa) w(\kappa') \right]$$
$$= \int [dw(\kappa)]_3 \exp\left[ -\frac{\pi}{2} \int_0^\infty d\kappa d\kappa' \frac{K_{-1}(\kappa, \kappa')}{\theta(\kappa)\theta(\kappa')} w(\kappa) w(\kappa') \right]. \tag{3.9}$$

These expressions can all be derived by translating from the corresponding measures in the mode basis. The kernel $K_{-1}(k, k')$ is given by the sum $\sum_{n=1}^{\infty} (2n-1)^{-1} v_{2n-1}(\kappa) v_{2n-1}(\kappa')$. An explicit formula for $K_{-1}$ was found in [33, 34], but we will not need it.

It is instructive to relate eq.(3.7) to other half-string bases. A simple computation shows,

$$l(\kappa) = \sqrt{2} \sum_{n=1}^{\infty} \left( \sqrt{2n} v_{2n}(\kappa) x_{2n} - \sqrt{2n-1} v_{2n-1}(\kappa) \tanh \frac{\pi\kappa}{4} x_{2n-1} \right)$$
$$= \sqrt{2} \sum_{n=1}^{\infty} \sqrt{2n} v_{2n}(\kappa) \left( x_{2n} + \sum_{m=1}^{\infty} T_{2n,2m-1} x_{2m-1} \right)$$
$$= \sqrt{2} \sum_{n=1}^{\infty} \sqrt{2n} v_{2n}(\kappa) l_{2n} = -\tanh \frac{\pi\kappa}{4} \sqrt{2} \sum_{n=1}^{\infty} \sqrt{2n-1} v_{2n-1}(\kappa) l_{2n-1},$$

where $l_{2n}$ denotes even cosine half-string Fourier modes and $l_{2n-1}$ denotes odd cosine half-string Fourier modes. Another simple computation can relate $l(\kappa)$ to the left half of the string, $l(\sigma) = x(\sigma)$, $\sigma \in [0, \frac{\pi}{2})$:

$$l(\kappa) = -\frac{2}{\sqrt{\pi}} \frac{\sinh \frac{\pi\kappa}{4}}{N(\kappa)} \int_0^{\pi/2} d\sigma \sec \sigma M(\kappa, \sigma) l(\sigma)$$
$$\bar{l}(\kappa) = -\frac{2}{\sqrt{\pi}} \frac{\cosh \frac{\pi\kappa}{4}}{N(\kappa)} \int_0^{\pi/2} d\sigma \sec \sigma M(\kappa, \sigma) l(\sigma), \quad (3.10)$$

where $M(\kappa, \sigma)$ can be derived by using the generating function eq.(3.2) to evaluate the sum,

$$M(\kappa, \sigma) = \frac{2N(\kappa) \cos \sigma}{\sqrt{\pi} \cosh \frac{\pi\kappa}{4}} \sum_{n=1}^{\infty} \sqrt{2n-1} v_{2n-1}(\kappa) \cos(2n-1)\sigma$$
$$= \frac{1}{\sqrt{\pi}} \cos\left(\frac{\kappa}{2} \tanh^{-1} \sin \sigma\right). \quad (3.11)$$

The $M$'s obey orthogonality relations:

$$\int_0^{\pi/2} d\sigma \sec \sigma M(\kappa, \sigma) M(\kappa', \sigma) = \delta(\kappa - \kappa')$$
$$\int_0^{\infty} d\kappa M(\kappa, \sigma) M(\kappa, \sigma') = \cos \sigma \delta(\sigma - \sigma'). \quad (3.12)$$

At first sight, the functions $M(\kappa, \sigma)$ seem quite exotic. It is reasonable to describe half string configurations using a basis of cosines, as is done in other treatments[11, 13]; why would anyone describe half strings in a basis of $M(\kappa, \sigma)$s? Actually, the $M$'s are special: they are eigenfunctions of the midpoint preserving reparameterization generator,

$$\mathcal{K}_1 = L_1 + L_{-1} = -2i \cos \sigma \frac{\partial}{\partial \sigma}.$$

Specifically,

$$(\mathcal{K}_1)^2 M(\kappa, \sigma) = -\kappa^2 M(\kappa, \sigma).$$

This is not surprising, since the kappa basis is defined by the eigenvectors of the Neumann coefficients, which in turn were derived in ref.[25] from the observation that a matrix $K_1$, describing the action of $\mathcal{K}_1$ on $|V_3\rangle$, possessed the same eigenvectors as the Neumann coefficients.

$\mathcal{K}_1$ acts like a generator of translations, but translations whose length becomes increasingly shrunk towards the midpoint; so in a sense the midpoint appears infinitely "far away." Correspondingly, the eigenfunctions $M(\kappa, \sigma)$ look locally like sine waves, but their frequency is position-dependent and blows up towards $\sigma = \frac{\pi}{2}$. (See figure 1a). The situation is therefore quite different from the discrete Moyal formalism, which imposes very specific boundary conditions at the midpoint; the value of $M$ and its derivative at the midpoint is completely undefined. Actually, in the continuous basis we should think of the midpoint boundary condition in the sense of a singular Sturm-Liouville

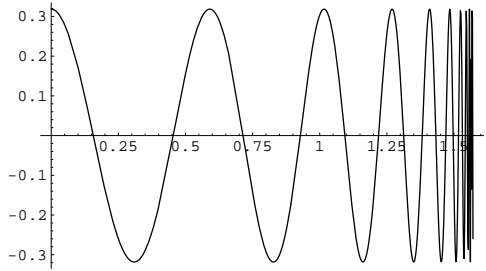

Figure1: An example, $M(10, \sigma)$.

system[8]; one should require that the eigenfunctions remain bounded as $\sigma$ goes to $\frac{\pi}{2}$. This effectively fixes $\kappa$ to be real.

It's also worth noticing that eq.(3.11) implies that a pure cosine wave in $\kappa$ space corresponds to a position eigenstate on the half-string. The frequency of the cosine wave and the position on the string are related: $\tanh \omega = \sin \sigma$. In particular, the ultraviolet in $\kappa$ space describes phenomena close to the string midpoint. Indeed, most singularities in the kappa basis manifest themselves in the apparent need to invoke functions whose frequency is formally infinite. We will see more of this in the following sections.

A simple but important question is how the variables $l(\kappa)$ and $\bar{l}(\kappa)$ represent the half string configuration $l(\sigma) = 1$ (or, equivalently, how $r(\kappa)$ and $\bar{r}(\kappa)$ represent $r(\sigma) = x(\pi - \sigma) = 1$). The relevance of this question should be explained. First, note that since we are working at zero momentum, we can for definiteness fix the string midpoint to be at zero: $x(\frac{\pi}{2}) = 0$. Then a string configuration with a kink discontinuity at the midpoint is,

$$x(\sigma) = \begin{cases} +1 & \text{for } \sigma \in [0, \frac{\pi}{2}) \\ 0 & \text{for } \sigma = \frac{\pi}{2} \\ -1 & \text{for } \sigma \in (\frac{\pi}{2}, \pi] \end{cases}.$$

Obviously, to describe this we need to use a basis capable of describing the function $l(\sigma) = 1$. There is reason to believe that such a configuration may be useful in string field theory, even though it looks somewhat singular; for instance it was shown in ref.[10] that the sliver functional, which is the canonical projector solution of VSFT, is independent of transformations which break a string at its midpoint. These considerations were further studied in ref.[25, 35, 36]. The discrete even Moyal basis, as mentioned before, cannot describe such phenomena. However, if we choose a half-string basis which *could* describe $l(\sigma) = 1$ we run into another problem: the basis accommodates configurations like:

$$x(\sigma) = \begin{cases} +1 & \text{for } \sigma \in [0, \frac{\pi}{2}) \\ 0 & \text{for } \sigma = \frac{\pi}{2} \\ +1 & \text{for } \sigma \in (\frac{\pi}{2}, \pi] \end{cases}. \tag{3.13}$$

---

[8]Perhaps it's amusing to mention that $M(\kappa, \sigma)$ would describe the vibrational modes of a string with linear mass density $\sec \sigma$ and tension $\cos \sigma$. Of course, the strings in string theory have mass density and tension equal to 1, in suitable units.

This pathological situation describes a string which has become estranged from its midpoint. There is no evidence that such configurations have any realization in string field theory, in fact there is some evidence to the contrary[25]. But now we seem to be stuck in a catch-22: if we cannot describe $l(\sigma) = 1$ we have too little freedom, but if we can we have too much[9]. This issue is at the heart of the difficulty with the split string approach to string field theory.

So back to the question: how does the continuous basis deal with this problem? It is simple to plug $l(\sigma) = 1$ into eq.(3.10),

$$\bar{l}(\kappa) = -\frac{2}{\sqrt{\pi}} \frac{\cosh\frac{\pi\kappa}{4}}{N(\kappa)} \int_0^{\pi/2} d\sigma \sec\sigma M(\kappa, \sigma) = -\frac{2}{\pi}\delta(\kappa)$$
$$l(\kappa) = -\frac{2}{\sqrt{\pi}} \frac{\sinh\frac{\pi\kappa}{4}}{N(\kappa)} \int_0^{\pi/2} d\sigma \sec\sigma M(\kappa, \sigma) = -\frac{2}{\pi}\sinh\frac{\pi\kappa}{4}\delta(\kappa) = 0.$$

The situation is similar to that in the discrete basis: In one basis you can represent $l(\sigma) = 1$ but in the other you can't. However, the crucial point is that neither of these half string bases is really equivalent to the continuous Moyal product described in eq.(3.5). The reason is that, in order to construct a split string description, we implicitly had to make a change of basis which normalized the noncommutativity parameter to one. There are two ways to do this, corresponding to the two half string formalisms:

$$\tilde{x}(\kappa) = \frac{x(\kappa)}{\theta(\kappa)} \qquad [\tilde{x}(\kappa), y(\kappa')]_\star = 2i\delta(\kappa - \kappa') \Longrightarrow \bar{l}(\kappa), \bar{r}(\kappa)$$
$$\tilde{y}(\kappa) = \frac{y(\kappa)}{\theta(\kappa)} \qquad [x(\kappa), \tilde{y}(\kappa')]_\star = 2i\delta(\kappa - \kappa') \Longrightarrow l(\kappa), r(\kappa).$$

The problem is that $\theta(\kappa)$ possesses a zero mode, $\theta(\kappa)\delta(\kappa) = 0$, so we cannot really invert it to make this change of basis. The zero mode of $\theta(\kappa)$ is the twin brother of the zero mode of $T$, and as we can see it causes the same problems.

This clearly explains why the noncommutativity parameter in the kappa basis vanishes continuously towards $\kappa = 0$: If it did not, the basis would either have too many or too few degrees of freedom to describe a reasonable string. This also sheds some light on why the Moyal formulation of the star algebra is slightly preferable to the matrix/split string approach. With a Moyal product, we have a deformation parameter $\theta$ which can be taken to zero in a meaningful way to recover a purely commutative algebra. This is not true for a matrix algebra. While the matrix and Moyal descriptions are indeed isomorphic for $\theta \neq 0$, they are not equivalent when $\theta = 0$ and this is why the split string formalism in any basis always has too many or too few degrees of freedom.

In ref.[25] they showed that the Neumann matrix $M = CV^{11}$ (using standard notation) has a doubly degenerate spectrum of eigenvectors, $v_{2n}(\kappa)$ and $v_{2n-1}(\kappa)$, for $\kappa$ strictly greater than zero; for $\kappa = 0$, however, $M$ only has *one* twist odd eigenvector, with eigenvalue $-\frac{1}{3}$:

$$v_{2n-1}(0) = \frac{1}{\sqrt{\pi}} \frac{(-1)^{n+1}}{\sqrt{2n-1}}.$$

---
[9]Another interpretation of eq.(3.13) advocated in ref.[13] is that it really represents a string which has been translated to $x(\sigma) = 1$, and $x(\pi/2) = 0$ is a residual artifact. So in this perspective the unwanted degree of freedom can be used to describe strings at nonzero momentum. This is an interesting point of view, but it seems better to represent nonzero momentum directly with string zero modes, rather than as a singular limit of the higher mode excitations.

There *is* a twist even counterpart to this, which after normalizing by a vanishing factor of $\kappa$ is,

$$v'_{2n}(0) = \frac{1}{\sqrt{\pi}} \frac{(-1)^{n+1}}{\sqrt{2n}} \left(1 + \frac{1}{3} + \frac{1}{5} + ... + \frac{1}{2n-1}\right), \tag{3.14}$$

but in ref.[25] they gave several reasons for excluding this vector from the spectrum of $M$. Our analysis provides a more geometrical justification for the exclusion of this eigenvector, since the corresponding Moyal coordinate,

$$x'(0) = \lim_{\kappa \to 0} \frac{x(\kappa)}{\kappa} = \sqrt{2} \sum_{n=1}^{\infty} \sqrt{2n} v'_{2n}(0) x_{2n}$$
$$= -\frac{1}{4\sqrt{\pi}} \int_0^\pi d\sigma |\sec \sigma| x(\sigma), \tag{3.15}$$

is solely responsible for describing the isolated midpoint string configuration, eq.(3.13):

$$\frac{x(\kappa)}{\theta(\kappa)} = \tfrac{1}{2}\left[\bar{l}(\kappa) + \bar{r}(\kappa)\right] = -\frac{2}{\pi} \delta(\kappa) = \frac{4}{\pi} x'(\kappa).$$

Since we do not expect such a configuration to play a role in the theory, it seems reasonable to exclude $v'_{2n}(0)$ from the spectrum of $M$. Accordingly, all "well-behaved" string fields should be independent of $x'(0)$:

$$\frac{\delta \Psi}{\delta x'(0)} = \lim_{\kappa \to 0} \kappa \frac{\delta \Psi}{\delta x(\kappa)} = 0. \tag{3.16}$$

Since $\delta/\delta x'(0)$ is a derivation, this constraint is preserved by star multiplication and so is consistent with the structure of the theory.

## 4. Philosophy of associativity anomalies

Having elucidated the star product and its midpoint structure in some detail, we now consider an interesting phenomenon in the algebra of string fields: the breakdown of associativity. These "associativity anomalies" have appeared many times in the literature down through the years (see, for example, [3, 4, 37, 9]), and have long been a source of puzzlement and speculation. We will study a particularly simple example concerning the product,

$$\bar{x} * \Psi * P_L \tag{4.1}$$

where $\Psi$ is a sufficiently well-behaved string functional (such as one in the perturbative Fock space), $P_L$ is the momentum of a half string, and $\bar{x}$ is the "center of mass" position relative to the midpoint (which by convention we set to vanish, $x(\frac{\pi}{2}) \equiv 0$). Explicitly,

$$P_L = \frac{\sqrt{2}}{\pi} \sum_{n=1}^{\infty} \frac{(-1)^{n+1}}{2n-1} p_{2n-1} = \int_0^{\pi/2} d\sigma p(\sigma) = -\frac{1}{\sqrt{\pi}} y(0)$$

$$\bar{x} = \sqrt{2} \sum_{n=1}^{\infty} (-1)^{n+1} x_{2n} = \frac{1}{\pi} \int_0^{\pi} d\sigma x(\sigma) = -2 \int_0^{\infty} \frac{d\kappa}{\kappa N(\kappa)} x(\kappa). \qquad (4.2)$$

Eq.4.1 closely resembles a similar product studied many ago by Horowitz and Strominger[3], and may be considered a canonical example of the associativity anomaly. We revisit this product in light of our more refined understanding, finding, indeed, that the product eq.4.1 in general fails to be associative, the failure being directly related to ambiguous products of $T, R$, and $v$ discussed in Section 2. Curiously, however, we will find that the loss of associativity in general depends quite sensitively on how the star products in eq.4.1 are evaluated. It is therefore not quite correct to speak of the ambiguity of eq.4.1 as an anomaly in associativity, since parentheses in themselves do not seem to be enough to unambiguously define the product.

The problem with evaluating the product eq.4.1 turns out to be directly related to the commutator,

$$[\bar{x}, P_L]_\star,$$

If we attempt evaluate this expression, say in the mixed basis, we immediately run into an ambiguous double sum,

$$i \frac{4}{\pi} \sum_{m,n=1}^{\infty} (-1)^{m+1} T_{2m,2n-1} \frac{(-1)^{n+1}}{2n-1}. \qquad (4.3)$$

As observed numerous times[13, 9, 15], the value of this sum depends on the order in which the summation is carried out. We can see the problem in the kappa basis too[16]:

$$[\bar{x}, P_L]_\star = \frac{2}{\sqrt{\pi}} \int_0^{\infty} \frac{d\kappa}{\kappa N(\kappa)} 2i \tanh \frac{\pi\kappa}{4} \delta(\kappa).$$

Again, this integral is ill-defined, since the integrand has delta function support on the boundary of the region of integration. It is probably meaningless to talk about the value of this commutator without specifying a prescription for regulating these expressions. One approach would be to impose a mode number cutoff $N$ to $P_L$ and $N'$ to $\bar{x}$, and then take $N, N'$ to infinity in a way which specifies an order for evaluating the double sum:

$$\lim_{N' \to \infty} \lim_{N \to \infty} [\bar{x}(N'), P_L(N)]_\star = iwTv = 0$$
$$\lim_{N \to \infty} \lim_{N' \to \infty} [\bar{x}(N'), P_L(N)]_\star = iv\bar{T}w = ivv = i. \qquad (4.4)$$

Here, in the notation of ref.[9], we introduced an even moded vector $w_{2n} = \sqrt{2}(-1)^{n+1} \in \mathcal{H}^*_{even}$ satisfying $\bar{T}w = v$. The ambiguity above comes about because $w$ and $v$ are not in $\mathcal{H}'_{even}$ and $\mathcal{H}_{odd}$ respectively.

We could similarly try to impose the mode number cutoff in the kappa basis, but this would result in somewhat unpleasant formulas. To make sense of $[\bar{x}, P_L]_\star$ in this context we take a more refined point of view. The key point is to notice that coordinates of the form

$$\int_0^\infty d\kappa \left[f(\kappa)y(\kappa) + g(\kappa)x(\kappa)\right],$$

only have a well-defined Moyal-star commutator if $f(\kappa)$ and $g(\kappa)$ live in a sufficiently smooth space of test functions, since calculating the commutator inevitably involves integrating $f$ and $g$ against a delta function. The problem with calculating $[P_L, \bar{x}]$ is that $f$ and $g$ in this case are not smooth functions; they are distributions. We can see this explicitly by regulating the formulas for $P_L$ and $\bar{x}$ as follows:

$$P_L(\omega) = \frac{\sqrt{2}}{\pi i} \sum_{n=1}^\infty \frac{(i\tanh\omega)^{2n-1}}{2n-1} p_{2n-1} = -\frac{1}{\pi} \int_{-\infty}^\infty \frac{d\kappa}{\kappa N(\kappa)} \sin\omega\kappa\, y(\kappa)$$

$$\bar{x}(\omega) = -\sqrt{2} \sum_{n=1}^\infty (i\tanh\omega)^{2n} x_{2n} = -2 \int_0^\infty \frac{d\kappa}{\kappa N(\kappa)} [1 - \cos\omega\kappa] x(\kappa). \qquad (4.5)$$

Taking $\omega$ to infinity we should recover $P_L$ and $\bar{x}$, but the coefficients in the integrand oscillate with infinite frequency and diverge. What should we make of this? The rapid oscillation of $\sin\omega\kappa$ and $1 - \cos\omega\kappa$ is not necessarily a problem if these singular objects always appear integrated against a nicely behaved test function; then the regulated formulas eq.(4.5) converge to eq.(4.2) in the sense of distributions: $\sin\omega\kappa/\kappa$ converges to a delta function, while $(1 - \cos\omega\kappa)/\kappa$ converges to the principal value. Apparently, calculating $[\bar{x}, P_L]_\star$ requires us to integrate over a product of distributions, which is mathematical nonsense. Still, we can certainly calculate $[\bar{x}(\omega), P_L(\omega')]_\star$, for finite $\omega$ and $\omega'$. Taking $\omega$ and $\omega'$ to infinity in different orders,

$$\lim_{\omega' \to \infty} \lim_{\omega \to \infty} [\bar{x}(\omega'), P_L(\omega)]_\star = 2i \lim_{\omega' \to \infty} \frac{2}{\sqrt{\pi}} \int_0^\infty d\kappa \frac{\tanh\frac{\pi\kappa}{4}}{\kappa N(\kappa)} [1 - \cos\omega'\kappa]\delta(\kappa) = 0$$

$$\lim_{\omega \to \infty} \lim_{\omega' \to \infty} [\bar{x}(\omega'), P_L(\omega)]_\star = 2i \lim_{\omega \to \infty} \frac{2}{\pi} \int_{-\infty}^\infty d\kappa \frac{\tanh\frac{\pi\kappa}{4}}{2\kappa \sinh\frac{\pi\kappa}{2}} \sin\omega\kappa = i.$$

we find the same answers as we did imposing a mode cutoff in the mixed basis eq.(4.4). Note that this discussion is related to our earlier observation that the extreme ultraviolet in the $\kappa$ basis is related to midpoint subtleties.

Having seen that $\bar{x}$ and $P_L$ are somewhat singular, it is natural to define star products with these guys via their regulated formulas. Thus,

$$\Psi * P_L \equiv \lim_{\rho \to \infty} \Psi * P_L(\rho)$$
$$\bar{x} * \Psi \equiv \lim_{\rho \to \infty} \bar{x}(\rho) * \Psi, \qquad (4.6)$$

where $\rho$ is a convenient regulator. Armed with this definition, we return to the question of evaluating the tricky product $\bar{x} * \Psi * P_L$. The calculation is simplified by noting,

$$\bar{x}(\rho) \star A = \bar{x}(\rho)A + \tfrac{1}{2}[\bar{x}(\rho), A]_\star$$
$$A \star P_L(\rho) = AP_L(\rho) + \tfrac{1}{2}[A, P_L(\rho)]_\star.$$

Multiplying and writing out the terms,

$$\begin{aligned}
\bar{x}(\rho) \star \Psi \star P_L(\rho') &= \bar{x}(\rho)\Psi P_L(\rho') + \tfrac{1}{2}[\bar{x}(\rho), \Psi]_\star P_L(\rho') \\
&\quad + \tfrac{1}{2}\bar{x}(\rho)[\Psi, P_L(\rho')]_\star + \tfrac{1}{4}[\bar{x}(\rho), [\Psi, P_L(\rho')]_\star]_\star \\
&\quad + \tfrac{1}{2}[\bar{x}(\rho), P_L(\rho')]_\star \Psi.
\end{aligned} \quad (4.7)$$

We can now calculate the associator, cancelling terms which are manifestly equal regardless of the order of limits:

$$(\bar{x} \star \Psi) \star P_L - \bar{x} \star (\Psi \star P_L) =$$
$$\tfrac{1}{2}\left[\lim_{\rho'\to\infty}\lim_{\rho\to\infty} - \lim_{\rho\to\infty}\lim_{\rho'\to\infty}\right]\left(\tfrac{1}{2}[\bar{x}(\rho), [\Psi, P_L(\rho')]_\star]_\star + [\bar{x}(\rho), P_L(\rho')]_\star \Psi\right).$$

For a well behaved string functional we can argue that the first term on the right hand side vanishes. For example, in the mixed basis this term looks like,

$$[\bar{x}(N), [\Psi, P_L(N')]_\star]_\star = \frac{2}{\pi}\sum_{n=1}^{N}\sum_{n'=1}^{N'}\sum_{k,l=1}^{\infty} \frac{(-1)^{n'+1}(-1)^{n+1}}{2n'-1} T_{2n,2k-1} T_{2l,2n'-1} \frac{\partial^2 \Psi}{\partial p_{2k-1}\partial x_{2l}}. \quad (4.8)$$

Depending on how the mixed partial derivative evaluates, this expression might have an ambiguous $N, N' \to \infty$ limit. However, for the ground state wavefunctional,

$$\frac{\partial^2 \Psi_{|0\rangle}}{\partial p_{2k-1}\partial x_{2l}} = 4\frac{2l}{2k-1}p_{2k-1}x_{2l}\Psi_{|0\rangle},$$

and noting that $\Psi_{|0\rangle}$ only has support on configurations satisfying

$$\sum_{n=1}^{\infty}\frac{p_{2k-1}^2}{2k-1} < \infty \qquad \sum_{n=1}^{\infty} 2l x_{2l}^2 < \infty,$$

we can see that the $N, N' \to \infty$ of eq.(4.8) limit is unambiguous (in fact, it is zero). Therefore, the ambiguity of $[\bar{x}, P_L]_\star$ gives us the only contribution to the associator,

$$(\bar{x} \star \Psi) \star P_L - \bar{x} \star (\Psi \star P_L) = \tfrac{i}{2}\Psi. \quad (4.9)$$

Hence associativity is anomalous. Note that this result is independent of our chosen regulator and basis on $\mathbb{P}$ (it appears even in the kappa basis, contrary to the speculation of ref.[17]).

What has gone awry here? How can we reconcile this non-associativity with the "matrix product" intuition for the algebra of string fields? To gain more insight it would seem useful to attempt this calculation in position space, rather than the phase space $\mathbb{P}$ of even coordinates and odd momenta, since this way we can draw more usefully from our geometrical intuition. Unfortunately, the calculation in position space only makes matters more confusing, since the anomaly we found in eq.4.9 actually no longer appears! Let us see how this happens, and why. In position space, (regulated) multiplication by $\bar{x}$ and $P_L$ appears as,

$$\bar{x}(\rho) \star A = \frac{2}{\pi} \int_0^{\pi/2} d\sigma f_\rho(\sigma) x(\sigma) A[x(\sigma)]$$
$$A \star P_L(\rho) = i \int_{\pi/2}^{\pi} d\sigma \, g_\rho(\sigma) \frac{\delta}{\delta x(\sigma)} A[x(\sigma)]. \qquad (4.10)$$

The explicit form of $f_\rho$ and $g_\rho$ depends on the chosen regulator, but at any rate when $\rho \to \infty$ $f$ and $g$ become equal to one (see figures 2,3). Note also that these two integrals range over opposite halves of the string; calculating $\bar{x}(\rho) * \Psi * P_L(\rho')$ we can see that this is required by associativity of the star product. However, at $\rho = \infty$ a possible subtlety arises: $A \star P_L$ can be written as an integral of $i\delta/\delta x(\sigma) A$ over *both* halves of the string, with opposite sign:

$$A \star P_L = i \int_{\pi/2}^{\pi} d\sigma \, \frac{\delta}{\delta x(\sigma)} A[x(\sigma)]$$
$$= -\frac{i}{2} \left[ \int_0^{\pi/2} d\sigma \, \frac{\delta}{\delta x(\sigma)} - \int_{\pi/2}^{\pi} d\sigma \, \frac{\delta}{\delta x(\sigma)} \right] A[x(\sigma)].$$

This is possible because $A \star P_L$ describes a differential where the right half of the string is broken away from the midpoint, which because we are working at zero momentum is equivalent to breaking the left and right halves symmetrically away from each other (see figure 2). Keeping this in mind, we can calculate,

$$\bar{x} \star (\Psi \star P_L) = \lim_{\rho \to \infty} \frac{2i}{\pi} \int_0^{\pi/2} d\sigma f_\rho(\sigma) x(\sigma) \left[ \int_0^{\pi/2} d\sigma \, \frac{\delta}{\delta x(\sigma)} - \int_{\pi/2}^{\pi} d\sigma \, \frac{\delta}{\delta x(\sigma)} \right] \Psi[x(\sigma)]$$
$$= -\frac{i}{\pi} \int_0^{\pi/2} d\sigma x(\sigma) \left[ \int_0^{\pi/2} d\sigma \, \frac{\delta}{\delta x(\sigma)} - \int_{\pi/2}^{\pi} d\sigma \, \frac{\delta}{\delta x(\sigma)} \right] \Psi[x(\sigma)].$$

Now consider the opposite product, $(\bar{x} \star \Psi) \star P_L$. According to eq.(4.10), $\bar{x}(\rho) \star A$ multiplies $A$ by the average value of $x(\sigma)$ on the left half of the string weighted by $f_\rho(\sigma)$ (see figure 3). For finite values of $\rho$ the integral $\int_0^{\pi/2} d\sigma f_\rho$ vanishes— this indeed must be the case if multiplication by $P_L$ is to be associative: $(\bar{x}(\rho) * \Psi) * P_L = \bar{x}(\rho) * (\Psi * P_L)$. However, precisely at $\rho = \infty$ the integral $\int_0^{\pi/2} d\sigma f_\infty$ ($= \int_0^{\pi/2} d\sigma 1$) does *not* vanish. Viewing figure 3, we see this occurs because an increasingly narrow and deep "well" in $f_\rho$ gets pushed off of the region of integration towards the midpoint, and finally at $\rho = \infty$ the well disappears completely. When this happens, the quantity $\bar{x} \star \Psi$ suddenly depends on the spurious midpoint isolating coordinate $x'(0)$ discussed in section 3. The dependence on $x'(0)$ means that when multiplying $\bar{x} \star \Psi$ by $P_L$ we cannot apply our previous argument that $P_L$ acts nontrivially on both halves of the string. In fact,

$$(\bar{x} \star \Psi) \star P_L = \lim_{\rho \to \infty} \frac{2i}{\pi} \int_{\pi/2}^{\pi} d\sigma \, g_\rho(\sigma) \frac{\delta}{\delta x(\sigma)} \int_0^{\pi/2} d\sigma x(\sigma) \Psi[x(\sigma)]$$
$$= -\frac{i}{\pi} \int_0^{\pi/2} d\sigma x(\sigma) \left[ \int_0^{\pi/2} d\sigma \, \frac{\delta}{\delta x(\sigma)} - \int_{\pi/2}^{\pi} d\sigma \, \frac{\delta}{\delta x(\sigma)} \right] \Psi[x(\sigma)].$$

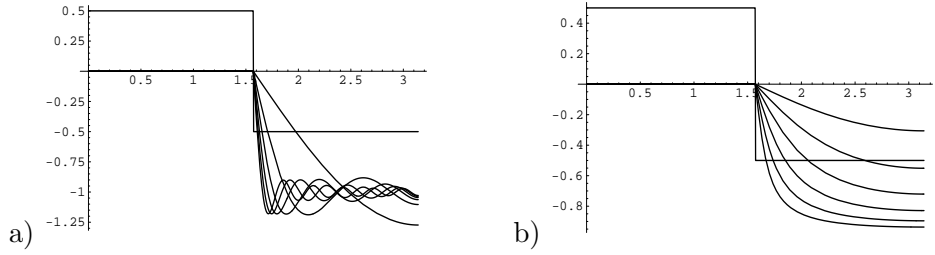

Figure 2: Graphs of $g_\rho$ as a function of the regulator in a) the discrete/Mixed basis for $N = 1, 3, 5, 7, 9, 11$ and b) in the continuous basis for $4\omega = 1, 2, 3, 4, 5, 6$.

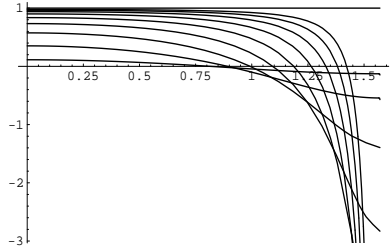

Figure 3: $f_\rho$ in the continuous basis, for $\rho = 2\omega = 1, ..., 9$. Note that for $\omega < \infty$ the area under the curve vanishes.

Therefore there is no associativity anomaly. Clearly, with all of these delicate midpoint issues an anomaly had the chance to appear, but it happens that in position space the subtleties conspire to cancel out.

How do we reconcile this with our previous result? The problem is that the limit $\rho \to \infty$ does not commute with the operation of taking the Fourier transform. Translating eq.(4.7) into position space, we see that the offending term is,

$$\tfrac{1}{2}[\bar{x}(\rho), \Psi]_\star P_L(\rho') = -\frac{i}{2\pi} \int_0^\pi g_{\rho'}(\sigma) \frac{\delta}{\delta x(\sigma)} \left[ \int_0^\pi d\sigma f_\rho(\sigma) x(\sigma) \right] \Psi,$$

where in our notation $f_\rho$ and $g_\rho$ are odd functions around $\sigma = \frac{\pi}{2}$. In the phase space of even coordinates and odd momenta, the limits $\rho, \rho' \to \infty$ commute because,

$$\lim_{\rho \to \infty} \lim_{\rho' \to \infty} \tfrac{1}{2}[\bar{x}(\rho), \Psi]_\star P_L(\rho') = \tfrac{1}{2} \left( \lim_{\rho \to \infty} [\bar{x}(\rho), \Psi]_\star \right) \left( \lim_{\rho' \to \infty} P_L(\rho') \right).$$

However, in coordinate space, $P_L$ is a differential operator whose action on $[\bar{x}(\rho), \Psi]_\star$ needs to be evaluated before we take can take $\rho \to \infty$. Taking the derivative gives us an extra term,

$$-\frac{2i}{\pi} \int_0^{\pi/2} d\sigma f_\rho(\sigma) g_{\rho'}(\sigma) \Psi,$$

which has an ambiguous $\rho, \rho' \to \infty$ limit, cancelling the contribution to the associator from $[\bar{x}, P_L]_\star \Psi$.

To make matters even more complicated, note that the products $\bar{x} \star \Psi$ and $\Psi \star P_L$ need not be defined in regulated form through eq.(4.6). We could have taken the limit $\rho \to \infty$ before calculating the star product with $\Psi$, rather than after. Calculating in this order, we find:

$$\Psi \star P_L = \Psi P_L$$
$$\bar{x} \star \Psi = \bar{x}\Psi + i\frac{2\sqrt{2}}{\pi} \sum_{m=1}^{\infty} \frac{(-1)^{m+1}}{2m-1} \frac{\partial \Psi}{\partial p_{2m-1}}.$$

Then,

$$\bar{x} \star (\Psi \star P_L) = \bar{x}\Psi P_L + i\frac{2\sqrt{2}}{\pi} \sum_{m=1}^{\infty} \frac{(-1)^{m+1}}{2m-1} \frac{\partial \Psi}{\partial p_{2m-1}} + \frac{i}{2}\Psi$$
$$(\bar{x} \star \Psi) \star P_L = \bar{x}\Psi P_L + i\frac{2\sqrt{2}}{\pi} \sum_{m=1}^{\infty} \frac{(-1)^{m+1}}{2m-1} \frac{\partial \Psi}{\partial p_{2m-1}} P_L.$$

So,

$$(\bar{x} \star \Psi) \star P_L - \bar{x} \star (\Psi \star P_L) = -\tfrac{i}{2}\Psi.$$

Now the associator has *opposite* sign from eq.(4.9)! Apparently the limit $\rho \to \infty$ does not commute with star multiplication either.

It has been suggested many times[3, 4, 9] that the failure of associativity in the star algebra is a necessary feature indicating the emergence of closed string physics from open string field theory. Bars and Matsuo have recently argued that the anomaly is needed in VSFT too[15], in order to ensure that D-brane vacua have a nontrivial spectrum of fluctuations. These ideas are interesting, but they are based on the assumption that when associativity fails, it does so in a consistent and physically meaningful way. This seems not to be the case, at least from the perspective of our analysis. A more sober point of view is that associativity anomalies indicate that we are attempting to multiply objects which are outside the algebra of string fields, and really the star product of such objects should not be defined. It might be possible that one day physical considerations will dictate an unambiguous procedure for regulating and calculating the star product when associativity no longer holds, but at this point the definition of products like $\bar{x} \star \Psi \star P_L$ is rather a question of philosophy.

## 5. Ghosts

So far[10], our analysis has dealt with the matter sector exclusively. In bosonized form[13, 18], the star algebra of ghosts is identical[11] to the algebra in the matter sector, eq.(2.2). However, in the $bc$ ghost language, the algebra of string fields looks quite different from the matter sector because the ghost coordinates $b'(\sigma)$ and $c(\sigma)$ are identified using half-string

---

[10]This discussion is taken and extended from an earlier discussion in ref.[28].

[11]Restricting to the subalgebra of fields with ghost momentum $-3/2$ is equivalent to studying zero momentum in the matter sector

*anti*overlap conditions (overlap with a sign)[2, 28, 22]. This leads to some novel structures beyond what we've seen. Define the anticommuting ghost fields,

$$b(\sigma) = i\sqrt{2}\sum_{n=1}^{\infty} x_n \sin n\sigma \qquad x_n = \frac{i}{\sqrt{2}}(b_n - b_{-n})$$

$$\pi_c(\sigma) = \frac{1}{\pi}\left(b_0 + \sqrt{2}\sum_{n=1}^{\infty} q_n \cos n\sigma\right) \qquad q_n = \frac{1}{\sqrt{2}}(b_n + b_{-n})$$

$$c(\sigma) = c_0 + \sqrt{2}\sum_{n=1}^{\infty} y_n \cos n\sigma \qquad y_n = \frac{1}{\sqrt{2}}(c_n + c_{-n})$$

$$\pi_b(\sigma) = -\frac{i\sqrt{2}}{\pi}\sum_{n=1}^{\infty} p_n \sin n\sigma \qquad p_n = \frac{i}{\sqrt{2}}(c_n - c_{-n}).$$

$b_n$ and $c_n$ are the mode oscillators of the $bc$ ghost conformal field theory satisfying $\{b_n, c_{-m}\} = \delta_{m,n}$ (note that our use of the $x_n$ and $p_n$ differs from previous sections). The discussion is simplified by restricting to Siegel gauge[28], where we will choose a representation of the the string field as a functional of the eigenvalues of $u = \{x_{2n-1}, p_{2n}, y_{2n-1}, q_{2n}\}$ for $n \geq 1$. Just as in the matter sector, these eigenvalues are a set of coordinates a linear space $\mathbb{P}_{gh}$ with a real symmetric inner product,

$$\langle u, u' \rangle = i\sum_{n=1}^{\infty} \left(x_{2n-1}y'_{2n-1} - y_{2n-1}x'_{2n-1} - q_{2n}p'_{2n} + p_{2n}q'_{2n}\right)$$

We can then calculate the reduced star product[38] (= $b_0$ times the full star product) as:

$$\Psi *_{b_0} \Phi[u] = \mathcal{N}_{gh}\Psi \star \Phi[u],$$

where $\mathcal{N}_{gh}$ is a normalization[28] and $\star$ is a fermionic, non-anticommutative, "Moyal product" satisfying,

$$\{y_{2m-1}, q_{2n}\}_\star = 2R_{2m-1,2n} \quad \{x_{2m-1}, p_{2n}\}_\star = 2\frac{2m-1}{2n}R_{2m-1,2n}, \tag{5.1}$$

where $\{,\}_\star$ is the Moyal anticommutator and $R$ from eq.(2.6) plays the role of the non-anticommutativity parameter.

Just as in the matter sector, the star product is simplified by a unitary transformation on $\mathbb{P}_{gh}$ taking us into the kappa basis:

$$x_e(\kappa) = -\sqrt{2}\sum_{n=1}^{\infty} \frac{1}{\sqrt{2n}}v_{2n}(\kappa)q_{2n}$$

$$x_o(\kappa) = \sqrt{2}\sum_{n=1}^{\infty} \frac{1}{\sqrt{2n-1}}v_{2n-1}(\kappa)x_{2n-1}$$

$$y_e(\kappa) = \sqrt{2}\sum_{n=1}^{\infty} \sqrt{2n}v_{2n}(\kappa)p_{2n}$$

$$y_o(\kappa) = \sqrt{2}\sum_{n=1}^{\infty} \sqrt{2n-1}v_{2n-1}(\kappa)y_{2n-1}. \tag{5.2}$$

In this basis, the inner product on $\mathbb{P}_{gh}$ takes the form,

$$\langle u, u' \rangle = i \int_0^\infty d\kappa \left[ \vec{x}(\kappa) \cdot \sigma_z \cdot \vec{y}(\kappa)' - \vec{y}(\kappa) \cdot \sigma_z \cdot \vec{x}(\kappa)' \right]$$

where we defined $\vec{x} = (x_\text{e}, x_\text{o}), \vec{x} = (x_\text{e}, x_\text{o})$ and $\sigma_z$ is the $z$th Pauli matrix. The algebra eq.5.1 now takes the form,

$$\{\vec{x}(\kappa), \vec{y}(\kappa')\}_\star = 2i\sigma_y g(\kappa)\ \delta(\kappa - \kappa'), \tag{5.3}$$

where

$$g(\kappa) = \coth \tfrac{\pi\kappa}{4}.$$

Note, in particular, that $g(\kappa)$ has a pole at $\kappa = 0$, and no zeros.

As in the matter sector, it is clear that all of the interesting structure in this algebra lies at $\kappa = 0$. However, unlike the matter sector, the physical significance of this structure seems obscure. What does it mean that the $\kappa = 0$ coordinates have infinite non-anticommutativity? The odd Moyal coordinates at $\kappa = 0$,

$$x_\text{o}(0) = \sqrt{\tfrac{2}{\pi}} \left( \hat{x}_1 - \tfrac{1}{3}\hat{x}_3 + \tfrac{1}{5}\hat{x}_5 - ... \right) \quad y_\text{o}(0) = \sqrt{\tfrac{2}{\pi}} \left( \hat{y}_1 - \hat{y}_3 + \hat{y}_5 - ... \right),$$

do not correspond to any particularly simple string degree of freedom. Even more puzzling, the even coordinates *vanish* at $\kappa = 0$, yet somehow manage to have infinite non-anticommutativity with the odd coordinates! Surely $0 * x_\text{o}(0) = x_\text{o}(0) * 0 = 0$?

Part of the confusion here amounts to a simple problem of domains. The anticommutator eq.(5.2) must be thought of as an equation relating distributions on the appropriate space of test functions. In this case, the space should be bounded $C^\infty$ functions on the interval $\kappa \in [0, \infty)$ which vanish linearly at $\kappa = 0$: $\lim_{\kappa \to 0} |f(\kappa)/\kappa| < \infty$. On this space, define a pair of distributions,

$$g \cdot \delta(\kappa, \kappa') = g(\kappa')\bar{\delta}_\kappa(\kappa') = g(\kappa)\delta_{\kappa'}(\kappa)$$

so that,

$$\langle g \cdot \delta_\kappa,\ f \rangle = \int_0^\infty d\kappa' g \cdot \delta(\kappa', \kappa) f(\kappa') = \begin{cases} g(\kappa)f(\kappa) & \text{for } \kappa > 0 \\ \tfrac{2}{\pi} f'(0) & \text{for } \kappa = 0 \end{cases}$$

$$\langle g \cdot \bar{\delta}_\kappa,\ f \rangle = \int_0^\infty d\kappa' g \cdot \delta(\kappa, \kappa') f(\kappa') = \begin{cases} g(\kappa)f(\kappa) & \text{for } \kappa > 0 \\ 0 & \text{for } \kappa = 0 \end{cases}. \tag{5.4}$$

Both of these distributions are *almost* equivalent to $g(\kappa)\delta(\kappa - \kappa')$, but crucially differ at $\kappa, \kappa' = 0$; apparently, $g \cdot \delta(\kappa, \kappa')$ is not symmetric when considered as a distribution in $\kappa$ versus $\kappa'$. With these definitions, we should more precisely write the star algebra in the ghost sector as,

$$\{x_\text{e}(\kappa), y_\text{o}(\kappa')\}_\star = -\{y_\text{e}(\kappa), x_\text{o}(\kappa')\}_\star = 2g \cdot \delta(\kappa, \kappa') \tag{5.5}$$

with unwritten anti-commutators vanishing. The nontrivial component here is the value of the these expressions at $\kappa, \kappa' = 0$. Setting $\kappa = 0$ for the even coordinates, we see now that both sides of this equation vanish unambiguously as a result of the definition eq.5.4. To see that eq.(5.4) gives the right result for the odd coordinates at $\kappa' = 0$, let us use the mixed basis to calculate the anticommutator[12],

$$\{y_o(0), q_{2n}\}_\star = -2\sqrt{\frac{2}{\pi}} \sum_{m=1}^{\infty} (-1)^m R_{2m-1,2n} = -\frac{8\sqrt{2}}{\pi} \sqrt{2n}\, v'_{2n}(0).$$

On the other hand, using eq.(5.4) we have,

$$\begin{aligned} \{y_o(0), q_{2n}\}_\star &= -\sqrt{2}\sqrt{2n} \int_0^\infty d\kappa v_{2n}(\kappa)\{y_o(0), x_e(k)\}_\star \\ &= -\sqrt{2}\sqrt{2n} \int_0^\infty d\kappa v_{2n}(\kappa) g \cdot \delta(\kappa, 0) \\ &= -\frac{8\sqrt{2}}{\pi} \sqrt{2n}\, v'_{2n}(0). \end{aligned}$$

So the definition works out consistently. As an aside we mention that, in the matter sector, the correct space of test functions for the commutator eq.(3.5) is bounded $C^\infty$ functions on $\kappa \in [0, \infty)$. The zero in $\tanh \frac{\pi \kappa}{4}$ might allow one to extend to a space of test functions with a pole at the boundary, but as we've seen this has the effect of introducing the spurious coordinate $x'(0)$ into the theory.

Fixing domains brings a little more rigor into our formulae, but still the physical interpretation of the pole and the corresponding coordinates at $\kappa = 0$ remains unclear. Let us take a moment to clarify the situation. Writing $x_o(\kappa)$ and $y_o(\kappa)$ in terms of $c(\sigma)$ and $b(\sigma)$, we find

$$y_o(\kappa) = \frac{\cosh\frac{\pi\kappa}{4}}{\sqrt{\pi}N(\kappa)} \int_0^\pi d\sigma \sec\sigma M(\kappa,\sigma)\, c(\sigma)$$

$$x_o(\kappa) = \frac{\cosh\frac{\pi\kappa}{4}}{\sqrt{\pi}N(\kappa)} \int_0^\pi d\sigma \sec\sigma M(\kappa,\sigma)\, \tilde{b}(\sigma),$$

where,

$$\tilde{b}(\sigma) = \int_{\pi/2}^\sigma d\sigma' b(\sigma').$$

---

[12]This result may be derived from the following manipulation:

$$\begin{aligned} \{y_o(0), \sum_{n=1}^\infty z^{2n} q_{2n}\}_\star &= -2\sqrt{\frac{2}{\pi}} \sum_{m,n=1}^\infty (-1)^m z^{2n} R_{2m-1,2n} \\ &= \frac{4}{\pi}\sqrt{\frac{2}{\pi}} \int_0^{\pi/2} d\sigma(\sec\sigma) \left(\frac{2z^2}{(1+z^2)(1-z^2)} \frac{\cos^2\sigma}{1+(\frac{2z}{1-z^2}\sin\sigma)^2}\right) \\ &= \frac{4}{\pi}\sqrt{\frac{2}{\pi}} \frac{z}{1+z^2} \int_0^{\frac{2z}{1-z^2}} \frac{du}{1+u^2} \\ &= \frac{8}{\pi}\sqrt{\frac{2}{\pi}} \frac{z\tan^{-1}z}{1+z^2}. \end{aligned}$$

Taylor expanding in $z$ and recalling eq.(3.14) we get the answer.

With this formula it is easy to check that $x_o(0)$ and $y_o(0)$ are necessary to describe string configurations where $c(\sigma)$ and $\tilde{b}(\sigma)$ have a midpoint discontinuity. Suppose for example that $c(\sigma)$ is a step function which jumps from 1 to $-1$ at the midpoint. Then,

$$y_o(\kappa) = \frac{2\cosh\frac{\pi\kappa}{4}}{\sqrt{\pi}N(\kappa)}\int_0^{\pi/2} d\sigma \sec\sigma M(\kappa,\sigma) = \frac{2}{\sqrt{\pi}}\delta(\kappa),$$

so the configuration is described solely with $y_o(0)$. Another way of saying this is that variations with respect to $x_o(0)$ and $y_o(0)$ generate midpoint discontinuities,

$$\frac{\delta}{\delta y_o(0)} = \frac{\sqrt{\pi}}{2}\left(\int_0^{\pi/2}d\sigma\pi_c(\sigma) - \int_{\pi/2}^{\pi}d\sigma\pi_c(\sigma)\right) \equiv \sqrt{\pi}\Pi_{c,L}$$

$$\frac{\delta}{\delta x_o(0)} = \frac{\sqrt{\pi}}{2}\pi_b(\tfrac{\pi}{2}). \tag{5.6}$$

Apparently, the $\kappa = 0$ coordinates themselves do not have a simple interpretation in terms of string degrees of freedom, however *variations* with respect to them do: varying with respect to $x_o(0)$ gives the half-string momentum of the $c$ ghost, while varying with respect to $y_o(0)$ gives the $b$ ghost momentum at the midpoint.

To interpret the physical significance of the pole at $\kappa = 0$, this suggests that it may be more useful to understand the effect of the derivations $\delta/\delta x_o(0), \delta/\delta y_o(0)$ on the algebra than the properties of the $\kappa = 0$ coordinates themselves. In this vein, consider the product,

$$\Psi *_{b_0} \left(\frac{\delta}{\delta x_o(0)}\Phi\right)[u]$$

for two arbitrary, well-behaved string fields. We can calculate this using the kernel representation of the Moyal product[28]:

$$\Psi *_{b_0}\left(\frac{\delta}{\delta x_o(0)}\Phi\right)[u^3] = \mathcal{N}_{gh}\int [du^1][du^2]\Psi[u^1]\left(\frac{\delta}{\delta x_o^2(0)}\Phi[u^2]\right)K[u^1,u^2,u^3]$$

$$= (-1)^{G_\Phi+1}\mathcal{N}_{gh}\int [du^1][du^2]\Psi[u^1]\Phi[u^2]\frac{\delta}{\delta x_o^2(0)}K[u^1,u^2,u^3]$$

where in the last step we integrated by parts and $G_\Phi$ denotes the Grassmann parity of $\Phi$. To proceed we need the explicit formula for $K$:

$$K[u^1,u^2,u^3] = C\exp\left[-\frac{i}{2}\int_{-\infty}^\infty \frac{1}{g(\kappa)}\left(x_e^A(\kappa)K^{AB}y_o^B(\kappa) - x_o^A(\kappa)K^{AB}y_e^B(\kappa)\right)\right]$$

where $C$ is a constant (which we will not need) and $K^{AB}$ is a $3\times 3$ matrix,

$$K^{AB} = \begin{pmatrix} 0 & 1 & -1 \\ -1 & 0 & 1 \\ 1 & -1 & 0 \end{pmatrix}^{AB}$$

We find,

$$\frac{\delta}{\delta x_o^2(0)}K[u^1,u^2,u^3] = \lim_{\kappa\to 0}\frac{i}{2}\frac{1}{g(\kappa)}K^{2B}y_e^B(\kappa)K[u^1,u^2,u^3]$$

This clearly *vanishes* as a consequence of the pole in $g(\kappa)$ at $\kappa = 0$. The argument similarly goes through for $y_o(\kappa)$. This implies a surprising fact: any string field which is proportional to $\pi_b(\frac{\pi}{2})$ or $\Pi_{c,L}$ has vanishing star product with any other string field:

$$\Psi *_{b_0} \left[\pi_b(\tfrac{\pi}{2})\Phi\right] = 0, \quad \Psi *_{b_0} \left[\Pi_{c,L}(\tfrac{\pi}{2})\Phi\right] = 0 \tag{5.7}$$

Thus the interpretation of the pole at $\kappa = 0$ is clear: it implies the existence of null elements in the algebra of string fields. By cyclicity of the kernel we also have the relations,

$$\left[\pi_b(\tfrac{\pi}{2})\Psi\right] *_{b_0} \Phi = 0, \quad \left[\Pi_{c,L}(\tfrac{\pi}{2})\Psi\right] *_{b_0} \Phi = 0$$
$$\pi_b(\tfrac{\pi}{2})\left[\Psi *_{b_0} \Phi\right] = 0, \quad \Pi_{c,L}(\tfrac{\pi}{2})\left[\Psi *_{b_0} \Phi\right] = 0 \tag{5.8}$$

A very similar structure concerning $c(\frac{\pi}{2})$ and the full star product was discovered by Okuyama in ref.[31]. We have not reproduced his result, of course, since we are studying the reduced star product in Siegel gauge (also introduced in [31]). For comparison, we can rephrase the commutative nature of $P_L$ in the matter sector in an analogous form to equations 5.7-5.8:

$$P_L(\Psi * \Phi) = (P_L\Psi) * \Phi = \Psi * (P_L\Phi)$$

We have already seen that the vanishing of $\theta(\kappa)$ at $\kappa = 0$ provides an obstruction to realizing the open string star algebra using an operator/matrix product. Analogously, the pole in $g(\kappa)$ at $\kappa = 0$ can be interpreted as an obstruction to realizing the star algebra using a Moyal product. Clearly, a Moyal product with infinite noncommutativity parameter is a singular object, but more importantly any Moyal product trivially has an identity element, 1. However, as we have seen the pole in $g(\kappa)$ implies that the open string star product possesses null elements, so there cannot be a well-defined identity on all string fields. Indeed, it is not difficult to see that the product $1 \star \frac{\delta}{\delta x_o(0)} \Psi$ is not well-defined. Of course, subtleties with the identity string field are quite well-known, and our discussion explains why 1 usually, but not always, behaves like an identity for the open string star algebra.

## 6. Complex $\kappa$ and the midpoint

Consider the following pair of ghost coordinates:

$$b(\tfrac{\pi}{2}) = i\sqrt{2}\sum_{n=1}^{\infty}(-1)^{n+1}x_{2n-1}$$
$$c'(\tfrac{\pi}{2}) = \sqrt{2}\sum_{n=1}^{\infty}(2n-1)(-1)^{n+1}y_{2n-1}, \tag{6.1}$$

Interestingly, they are purely *anticommutative* under the open string product:

$$\{b(\tfrac{\pi}{2}),\ p_{2n}\}_\star = 0 \quad \{c'(\tfrac{\pi}{2}),\ q_{2n}\}_\star = 0.$$

This can be seen intuitively from the fact that the ghost overlap conditions identify the midpoint values of $b(\sigma)$ and $c'(\sigma)$ locally. More explicitly it can be seen from the fact that $\bar{R}$ has a zero mode, eq.(2.8).

However, this presents us with a puzzle. In the continuous basis, the non-anticommutativity parameter $g(\kappa) = \coth\frac{\pi\kappa}{4}$ has no zeros, so it seems impossible that any linear combination of $x_o(\kappa)$ and $y_o(\kappa)$ could anticommute with *all* $x_e(\kappa)$ and $y_e(\kappa)$. The unexpected twist here, however, is that $b(\frac{\pi}{2})$ and $c'(\frac{\pi}{2})$ are actually described by Moyal coordinates with an *imaginary argument*, and for imaginary $\kappa$, $\coth\frac{\pi\kappa}{4}$ has zeros. To explain this, it is convenient to define[13],

$$X_o(\kappa) = N(\kappa)x_o(\kappa) \quad Y_o(\kappa) = N(\kappa)y_o(\kappa).$$

This scaling is useful since $X_o$ and $Y_o$ are analytic everywhere on the complex plane, unlike $x_o$ and $y_o$; in particular, the change of basis eq.(5.2) becomes defined in terms of functions $N(\kappa)v_n(\kappa) = Ч_n(\kappa)/\sqrt{n}$, which as discussed in the appendix are just polynomials. So write $b(\frac{\pi}{2})$ in this basis:

$$b(\tfrac{\pi}{2}) = 2i\int_0^\infty d\kappa \left(\sum_{n=1}^\infty (-1)^{n+1} Ч_{2n-1}(\kappa)\right) \frac{X_o(\kappa)}{\frac{2}{\kappa}\sinh\frac{\pi\kappa}{2}}.$$

We can immediately see that something strange is going on, since the above series doesn't converge. However, the authors of ref.[33] suggested how this divergent sum could be formally defined. One writes,

$$\begin{aligned}b(\tfrac{\pi}{2}) &= 2i\int_0^\infty d\kappa \tfrac{1}{2}\sin\left(2\frac{d}{d\kappa}\right)\kappa \left[\sum_{n=1}^\infty \frac{(-1)^{n+1}}{2n-1} Ч_{2n-1}(\kappa)\right] \frac{X_o(\kappa)}{\frac{2}{\kappa}\sinh\frac{\pi\kappa}{2}} \\ &= 2i\sqrt{\pi}\int_0^\infty d\kappa \left[\tfrac{1}{2}\sin\left(2\frac{d}{d\kappa}\right)\kappa N(\kappa)\delta(\kappa)\right]\frac{X_o(\kappa)}{\frac{2}{\kappa}\sinh\frac{\pi\kappa}{2}}.\end{aligned} \quad (6.2)$$

The strange object in the brackets is some type of distribution—let us hold off on this a moment, and assume this makes sense. Integrating by parts,

$$b(\tfrac{\pi}{2}) = -\frac{i\sqrt{\pi}}{2}\int_{-\infty}^\infty d\kappa\,\delta(\kappa)N(\kappa)\kappa\Im\left[\frac{X_o(\kappa+2i)}{\frac{2}{\kappa+2i}\sinh\frac{\pi(\kappa+2i)}{2}}\right].$$

Cancelling the pole[14] at $\kappa = 0$ gives,

$$b(\tfrac{\pi}{2}) = iX_o(2i) \quad c'(\tfrac{\pi}{2}) = -Y_o(2i). \quad (6.3)$$

An identical calculation follows for $c'(\frac{\pi}{2})$. $X_o$ and $Y_o$ are evaluated exactly where $g(\kappa) = \coth\frac{\pi\kappa}{4}$ has a zero! Still this is very formal, since when calculating star products using these expressions one encounters delta functions evaluated with an *imaginary* argument. It is not at all clear what this means.

---

[13] We have seen now many times that attempting to "rescale" coordinates is dangerous, since it has the potential to alter the string's degrees of freedom in an unphysical way. However, for real $\kappa$ this particular rescaling is not a problem since $N(\kappa)$ is a smooth function without any zeros. For complex $\kappa$, $N(\kappa)$ has an infinite number of zeros (branch points), but these all cancel against corresponding singularities in the the coordinates. Therefore, we hold the view that the rescaled Moyal coordinates are really the ones we should have been using all along.

[14] A subtle point: One must be careful not to multiply $\kappa$ with $\delta(\kappa)$ before cancelling the pole, otherwise the answer would (wrongly) come out to be zero.

The conundrum is caused by the exotic distribution appearing in eq.(6.2). It is not a Schwartz distribution, since for a general $C^\infty$ function the infinite rank differential operator $\sin(2\frac{d}{dk})$ does not converge. The convergence of $\sin(2\frac{d}{dk})$ on a function $f(\kappa)$ is determined by the sum,

$$\sum_{n=0}^{\infty} \frac{f^{(2n+1)}(\kappa)}{(2n+1)!} 2^{2n+1} < \infty. \tag{6.4}$$

This says that a Taylor expansion of $f$ around any particular $\kappa$ must converge at least within a radius of 2; thinking of $f$ as a function in the complex plane, this means that $f$ must be analytic inside a band outside the real axis extending from $-2 < \Im(\kappa) < 2$. On this restricted space of test functions (much more restricted than $C^\infty$ functions), eq.(6.2) is a perfectly well defined distribution, and our formal calculation has rigorous meaning. To simplify notation, let us introduce the complex delta function distribution[33], defined:

$$\langle \delta_z, f \rangle = \int_{-\infty}^{\infty} d\kappa \, \delta(\kappa - z) f(\kappa) \equiv \left. \exp\left(z \frac{d}{d\kappa}\right) f(\kappa) \right|_{\kappa=0} \quad z \in \mathbb{C}. \tag{6.5}$$

$f$ must be analytic inside a band around the real axis $-a < \Im(\kappa) < a$ with $|z| < a$. Rewriting eq.(6.2) we get simply,

$$b(\tfrac{\pi}{2}) = i \langle \delta_{2i}, X_{\text{o}} \rangle \quad c'(\tfrac{\pi}{2}) = -\langle \delta_{2i}, Y_{\text{o}} \rangle.$$

This is really just a reexpression of eq.(6.3), but now we have a solid mathematical basis for calculating star products. For example,

$$\begin{aligned}
\{b(\tfrac{\pi}{2}), p_{2n}\}_\star &= \tfrac{1}{2} \int_{-\infty}^{\infty} d\kappa \, d\kappa' \, \delta(\kappa - 2i) \frac{Ч_{2n}(\kappa')}{\frac{2}{\kappa'} \sinh \frac{\pi\kappa'}{2}} \{X_{\text{o}}(\kappa), Y_{\text{e}}(\kappa')\}_\star \\
&= \int_{-\infty}^{\infty} d\kappa \, \delta(\kappa - 2i) Ч_{2n}(\kappa) \coth \tfrac{\pi\kappa}{4} \\
&= \left. \exp\left(2i \frac{d}{d\kappa}\right) Ч_{2n}(\kappa) \coth \tfrac{\pi\kappa}{4} \right|_{\kappa=0} = Ч_{2n}(2i) \cot \tfrac{\pi}{2} = 0.
\end{aligned}$$

This vanishes, as claimed, because the cotangent has a zero. Notice that $Ч_{2n}(\kappa) \coth \tfrac{\pi\kappa}{4}$ is analytic in the band $-2 \leq \Im(\kappa) \leq 2$ so the calculation is well defined (the pole of the cotangent gets cancelled by the zero of $Ч_{2n}$). An analogous calculation shows that $c'(\tfrac{\pi}{2})$ is anticommutative. The complex delta function eq.(6.5) is an example of a very general class of distributions studied in the mathematical literature, called "ultradistributions"[39].

For those who are uncomfortable with these formal computations, we can explicitly regulate and show that everything works fine. Consider,

$$b_\omega(\tfrac{\pi}{2}) = 2i \int_0^\infty d\kappa \frac{e^{2\omega} \cos \omega \kappa}{4N(\kappa)} x_{\text{o}}(\kappa) = i\sqrt{2} \sum_{n=1}^{\infty} \tfrac{1}{i}(i \tanh \omega)^{2n-1} x_{2n-1} + \mathcal{O}(e^{-\omega}). \tag{6.6}$$

We can rewrite this,

$$b_\omega(\tfrac{\pi}{2}) = \frac{i}{4} \int_{-\infty}^{\infty} d\kappa \frac{e^{i\omega(\kappa - 2i)}}{\frac{2}{\kappa} \sinh \frac{\pi\kappa}{2}} X_{\text{o}}(\kappa). \tag{6.7}$$

As $\omega \to \infty$, we in principle recover $b(\frac{\pi}{2})$, but the integrand oscillates with infinite frequency and diverging amplitude. Naively, it is surprising that this limit makes any sense at all, but with the proper restrictions it is perfectly sensible. Consider an almost everywhere analytic test function $f(\kappa)$ which has simple poles at $\kappa_n = a_n + ib_n$ and branch points at $\theta_n = c_n + id_n$. Evaluate the integral

$$\frac{1}{2\pi i}\int_{-\infty}^{\infty} d\kappa \frac{e^{i\omega(\kappa-2i)}}{\kappa-2i} f(\kappa) = f(2i) + \sum_{0<b_n<D} e^{-\omega(b_n-2)} e^{i\omega a_n} \frac{\mathrm{Res}(f,\kappa_n)}{\kappa_n - 2i}$$
$$+ \sum_{0<d_m<D} \frac{e^{\omega(2-d_m)}}{2\pi i}\int_{\theta_m} d\kappa \frac{e^{i\omega\Re(\kappa)}}{\kappa - 2i} f(\kappa) + \frac{e^{\omega(2-D)}}{2\pi i}\int_C d\kappa \frac{e^{i\omega\Re(\kappa)}}{\kappa - 2i} f(\kappa).$$

The contour $C$ runs from $-\infty$ to $\infty$ parallel and a distance $D > 2$ above the real axis. Branch cuts also have been chosen so that they run parallel to the real axis. Taking $\omega$ to infinity, depending on the location of the poles and branch points, we can have the following possibilities: If $0 \leq b_n < 2$ or $0 \leq d_n < 2$ the integral blows up; if $b_n, d_n = 2$ the integral oscillates but does not converge to a definite value; if $2 < b_n$ and $2 < d_n$ everything vanishes except $f(2i)$. Therefore:

$$\frac{1}{2\pi i}\lim_{\omega\to\infty}\int_{-\infty}^{\infty} d\kappa \frac{e^{i\omega(\kappa-2i)}}{\kappa-2i} f(\kappa) = \begin{cases} \text{undefined} & \text{if } f(\kappa) \text{ has poles or} \\ & \text{branch points for } 0 \leq \Im(\kappa) \leq 2 \\ f(2i) & \text{otherwise} \end{cases} \quad (6.8)$$

The sequence of functions $e^{i\omega(\kappa-2i)}/(\kappa-2i)$ therefore converges (in the sense of distributions) precisely to the complex delta function $\delta_{2i}$ discussed earlier[15]. We can use this result directly to calculate the commutator,

$$\{b(\tfrac{\pi}{2}), p_{2n}\}_\star = \frac{i}{2}\lim_{\omega\to\infty}\int_{-\infty}^{\infty} d\kappa \frac{e^{i\omega(\kappa-2i)}}{\frac{2}{\kappa}\sinh\frac{\pi\kappa}{2}} Ч_{2n}(\kappa) \coth\tfrac{\pi\kappa}{4},$$

which is seen to vanish as expected.

Since the cotangent has a zero at $\kappa = 2i$, the even Moyal coordinates should also be anticommutative at $\kappa = 2i$. The even coordinates have a simple interpretation in terms of the midpoint ghost momenta:

$$\langle \delta_{2i}, X_e \rangle = \tfrac{\pi}{i}\pi_c(\tfrac{\pi}{2}) \qquad \langle \delta_{2i}, Y_e \rangle = \pi \cdot \pi_b'(\tfrac{\pi}{2}).$$

Under the overlap $\pi_c(\tfrac{\pi}{2})$ and $\pi_b'(\tfrac{\pi}{2})$ are identified, and so indeed they should (naively) be anticommutative. However, translating to the mixed basis we see that the matrix $R$ does not have a corresponding zero mode. The vector which "should be" the zero mode, $w_{2n} = \sqrt{2}(-1)^{n+1}$, is far outside of the domain of $R$. How do we understand this? Calculating a typical anticommutator,

$$\{y_{2n-1}, \pi_c(\tfrac{\pi}{2})\}_\star = \frac{i}{\pi}\frac{2\sqrt{2}}{2n-1}\left\langle \delta_{2i}, Ч_{2n-1}(\kappa) \coth\tfrac{\pi\kappa}{4}\right\rangle = \infty. \quad (6.9)$$

---

[15]A delicate point: convergence of eq.(6.8) does not strictly require that $f$ be analytic for $0 \leq \Im(\kappa) \leq 2$. $f$ may have poles order $n \in \mathbb{Z} \geq 2$, and the integral will have the same value. By contrast, convergence of the differential operator $\exp(2i\frac{d}{d\kappa})$ requires analyticity everywhere in $0 \leq \Im(\kappa) \leq 2$, so eq.(6.8) and eq.(6.5) are not completely equivalent, though they agree whenever both expressions are defined.

This diverges because Ч$_{2n-1}(\kappa) \coth \frac{\pi\kappa}{2}$ has a pole at $\kappa = 0$, so it is not a valid test function for $\delta_{2i}$. (Note the importance of keeping track of domains—a naive evaluation would indeed give Ч$_{2n-1}(2i) \coth \frac{\pi \cdot 2i}{2} = 0$). However, the fact that eq.(6.9) diverges does not mean that $\pi_c(\frac{\pi}{2})$ is not anticommutative when multiplied with suitably restricted functionals—any functional of coordinates $\int_0^\infty f(\kappa) x_o(\kappa)$ for which $f$ is analytic in the band $-2 \leq \Im(\kappa) \leq 2$ and vanishes linearly (or faster) at $\kappa = 0$ will (anti)commute with $\pi_c(\frac{\pi}{2})$[16]. In particular, the Fourier modes $y_{2n-1}$ do not satisfy this restriction, and so their anticommutator with $\pi_c(\frac{\pi}{2})$ diverges.

Of course, the cotangent has zeros not just at $2i$, but for every odd multiple of $2i$—$\kappa = 6i, 10i, ...$. Therefore, on a suitably restricted space of functionals, the ghost sector actually has an infinite number of distinct anticommutative coordinates, four for every positive odd multiple of $2i$. Using some identities given in the appendix we find their explicit form to be:

$$\langle \delta_{2i(2n-1)}, X_e \rangle = \pi i \frac{(-1)^n}{2n-1} Ч_{2n-1}\left(2\frac{d}{d\sigma}\right) \pi_c(\sigma)\Big|_{\sigma=\frac{\pi}{2}}$$

$$\langle \delta_{2i(2n-1)}, Y_e \rangle = -\pi \frac{(-1)^n}{2n-1} Ч_{2n-1}\left(2\frac{d}{d\sigma}\right) \pi'_b(\sigma)\Big|_{\sigma=\frac{\pi}{2}}$$

$$\langle \delta_{2i(2n-1)}, X_o \rangle = i\frac{(-1)^n}{2n-1} Ч_{2n-1}\left(2\frac{d}{d\sigma}\right) b(\sigma)\Big|_{\sigma=\frac{\pi}{2}}$$

$$\langle \delta_{2i(2n-1)}, Y_o \rangle = \frac{(-1)^n}{2n-1} Ч_{2n-1}\left(2\frac{d}{d\sigma}\right) c'(\sigma)\Big|_{\sigma=\frac{\pi}{2}}. \quad (6.10)$$

As the zero gets further from the real axis, the corresponding anticommutative coordinates are given by progressively higher derivatives of the ghost coordinates and momenta at the string midpoint. Each coordinate is naively anticommutative according to the overlap conditions, but is in fact anticommutative only in an increasingly restricted sense as we move further up in the complex plane. This observation also holds in the matter sector, where the tangent has a zero for every even multiple of $2i$. For each zero there is a pair of commutative coordinates, ($X(\kappa) = N(\kappa)x(\kappa)$ and $Y(\kappa) = N(\kappa)y(\kappa)$ from eq.(3.1))

$$\langle \delta_{4in}, X \rangle = i\frac{(-1)^n}{2n} Ч_{2n}\left(2\frac{d}{d\sigma}\right) x'(\sigma)\Big|_{\sigma=\frac{\pi}{2}}$$

$$\langle \delta_{4in}, Y \rangle = -\pi \frac{(-1)^n}{2n} Ч_{2n}\left(2\frac{d}{d\sigma}\right) p(\sigma)\Big|_{\sigma=\frac{\pi}{2}}, \quad (6.11)$$

for $n \geq 1$.

The non(anti)commutativity parameters $\theta(\kappa), g(\kappa)$ also have an infinite sequence of poles in the complex plane, which as we have seen are equally important as zeros. Not surprisingly, variations with respect to the Moyal coordinates at these poles have a simple

---

[16]By our earlier discussion, functionals of these coordinates are independent of transformations which break the string at its midpoint.

interpretation in terms of midpoint degrees of freedom. In the ghost sector, we have

$$\left\langle \delta_{4in}, N \cdot \frac{\delta}{\delta x_{\mathrm{e}}} \right\rangle = \frac{i(-1)^n}{2n} Ч_{2n}\left(2\frac{\partial}{\partial \sigma}\right) c'(\sigma)\bigg|_{\sigma=\frac{\pi}{2}}$$

$$\left\langle \delta_{4in}, N \cdot \frac{\delta}{\delta x_{\mathrm{o}}} \right\rangle = \frac{i\pi(-1)^{n+1}}{2n} Ч_{2n}\left(2\frac{\partial}{\partial \sigma}\right) \pi'_b(\sigma)\bigg|_{\sigma=\frac{\pi}{2}}$$

$$\left\langle \delta_{4in}, N \cdot \frac{\delta}{\delta y_{\mathrm{e}}} \right\rangle = \frac{(-1)^n}{2n} Ч_{2n}\left(2\frac{\partial}{\partial \sigma}\right) b(\sigma)\bigg|_{\sigma=\frac{\pi}{2}}$$

$$\left\langle \delta_{4in}, N \cdot \frac{\delta}{\delta y_{\mathrm{o}}} \right\rangle = \frac{\pi(-1)^{n+1}}{2n} Ч_{2n}\left(2\frac{\partial}{\partial \sigma}\right) \pi_c(\sigma)\bigg|_{\sigma=\frac{\pi}{2}} \qquad (6.12)$$

while in the matter sector,

$$\left\langle \delta_{2i(2n-1)}, N \cdot \frac{\delta}{\delta x} \right\rangle = \frac{\pi(-1)^n}{2n-1} Ч_{2n-1}\left(2\frac{\partial}{\partial \sigma}\right) p(\sigma)\bigg|_{\sigma=\frac{\pi}{2}}$$

$$\left\langle \delta_{2i(2n-1)}, N \cdot \frac{\delta}{\delta y} \right\rangle = \frac{i(-1)^n}{2n-1} Ч_{2n-1}\left(2\frac{\partial}{\partial \sigma}\right) x'(\sigma)\bigg|_{\sigma=\frac{\pi}{2}} \qquad (6.13)$$

for $n \geq 1$. All of these operators, when acting on a string field, tend to make star products vanish. As before, when we consider poles increasingly far up in the complex plane, we encounter sequentially higher derivatives of string coordinates at the midpoint, and the operators can be considered null only in an increasingly restricted sense. It is worth noting the operators eq.6.12-6.13 naively annihilate the string vertex as a consequence of anti-overlap conditions. To see that this is the case, consider for example the operator,

$$\left\langle \delta_{2i}, N \cdot \frac{\delta}{\delta x} \right\rangle = -\pi p(\tfrac{\pi}{2})$$

By momentum conservation, the string momentum coordinate $p(\sigma)$ is identified in the vertex with anti-overlap:

$$p^A(\sigma)|V_3\rangle = -p^{A+1}(\pi - \sigma)|V_3\rangle \quad \sigma \in [0, \tfrac{\pi}{2}]$$

If we can assume this relation holds in the boundary case $\sigma = \tfrac{\pi}{2}$, we find,

$$p^1(\tfrac{\pi}{2})|V_3\rangle = -p^2(\tfrac{\pi}{2})|V_3\rangle = p^3(\tfrac{\pi}{2})|V_3\rangle = -p^1(\tfrac{\pi}{2})|V_3\rangle$$

implying that $p(\tfrac{\pi}{2})$ annihilates the vertex as claimed. A similar qualitative argument follows for all of the null operators eq.6.12-6.13.

As a word of admonition, we should realize that when dealing with with complex Moyal coordinates, ambiguities can creep in which are easy to miss. To see an example, consider the coordinate,

$$A = -\frac{i\sqrt{2}}{\pi} \sum_{n=1}^{\infty} \frac{(-1)^{n+1}}{2n} p_{2n}.$$

The anticommutator of $A$ with $b(\tfrac{\pi}{2})$ is ambiguous, as is easy to see in the mixed basis:

$$\{b(\tfrac{\pi}{2}),\ A\}_\star = \begin{cases} w' \cdot \bar{R}v' = w \cdot Tv = 0 \\ v' \cdot Rw' = v \cdot \bar{T}w = 1 \end{cases}, \tag{6.14}$$

where $w'_{2n} = \sqrt{2}\frac{(-1)^{n+1}}{(2n)^2} \in \mathcal{H}'^*_{\text{even}}$. This ambiguity is analogous to that of $[\bar{x}, P_L]_\star$ discussed earlier. We can find a regulated expression for $A$ in the continuous basis,[17]

$$\begin{aligned}
A_\omega &= \frac{i}{\pi}\int_{-\infty}^{\infty} \frac{d\kappa}{\kappa N(\kappa)}\left[\tfrac{1}{2}\psi(\tfrac{1}{2}+\tfrac{i\kappa}{4}) + \tfrac{1}{2}\psi(\tfrac{1}{2}-\tfrac{i\kappa}{4}) + \gamma + 2\ln 2\right.\\
&\qquad\qquad \left. -2e^{-2\omega}\frac{1+(\tfrac{\kappa}{2})^2 - \cos\omega\kappa + \tfrac{\kappa}{2}\sin\omega\kappa}{1+(\tfrac{\kappa}{2})^2}\right]y_{\text{e}}(\kappa)\\
&= \frac{i\sqrt{2}}{\pi}\sum_{n=1}^{\infty}\frac{(i\tanh\omega)^{2n}}{2n}p_{2n} + \mathcal{O}(e^{-4\omega}),
\end{aligned}$$

where $\psi(z) = \frac{d}{dz}\ln\Gamma(z)$ is the digamma function and $\gamma$ is Euler's constant. The curious thing about this regulated formula is that it converges as $\omega \to \infty$. However, when evaluating eq.(6.14) we cannot take the $\omega \to \infty$ limit too early, since the exponentially small oscillatory factor actually can make a finite contribution. We are lead to consider the integral,

$$\begin{aligned}
\{b_\omega(\tfrac{\pi}{2}), A_{\omega'}\}_\star &= \frac{1}{2\pi}\int_{-\infty}^{\infty} d\kappa \frac{\coth\tfrac{\pi\kappa}{4}}{\sinh\tfrac{\pi\kappa}{2}} e^{i\omega(\kappa-2i)}\left[\tfrac{1}{2}\psi(\tfrac{1}{2}+\tfrac{i\kappa}{4})+\tfrac{1}{2}\psi(\tfrac{1}{2}-\tfrac{i\kappa}{4})\right.\\
&\qquad \left. +\gamma+2\ln 2 - 2e^{-2\omega'}\frac{1+(\tfrac{\kappa}{2})^2-\cos\omega'\kappa+\tfrac{\kappa}{2}\sin\omega'\kappa}{1+(\tfrac{\kappa}{2})^2}\right].
\end{aligned}$$

Taking $\omega' \to \infty$ first, we must evaluate,

$$\{b_\omega(\tfrac{\pi}{2}), A\}_\star = \frac{1}{2\pi}\int_{-\infty}^{\infty}d\kappa \frac{e^{i\omega(\kappa-2i)}}{\sinh\tfrac{\pi\kappa}{2}}\left[\psi(\tfrac{1}{2}-\tfrac{i\kappa}{4})-\psi(\tfrac{1}{2})+\frac{\pi i}{4}\tanh\tfrac{\pi\kappa}{4}\right]\coth\tfrac{\pi\kappa}{4}.$$

Closing the contour in the upper half plane, and keeping in mind eq.(6.8), we see a potential problem with convergence from the double pole in $\frac{\coth\tfrac{\pi\kappa}{4}}{\sinh\tfrac{\pi\kappa}{2}}$ at $\kappa = 0$. However, this is cancelled against a second order zero in $\psi(\tfrac{1}{2}+\tfrac{i\kappa}{4}) + \psi(\tfrac{1}{2}-\tfrac{i\kappa}{4}) - \psi(\tfrac{1}{2})$. Furthermore, the zero of the cotangent cancels the pole in the tangent at $2i$, giving a nonzero contribution to the $\omega \to \infty$ limit,

$$\lim_{\omega\to\infty}\lim_{\omega'\to\infty}\{b_\omega(\tfrac{\pi}{2}), A_{\omega'}\}_\star = \frac{1}{2\pi}(2\pi i)(-\tfrac{1}{2})(2i) = 1,$$

---

[17]This equation follows from the following asymptotic formulas for large $\omega$, which we list for reference:

$$\sum_{n=1}^{\infty}\frac{(i\tanh\omega)^{2n}}{(2n)^{3/2}}v_{2n}(\kappa) \sim \frac{1}{\kappa N(\kappa)}\left[\tfrac{1}{2}\psi(\tfrac{1}{2}+\tfrac{i\kappa}{4})+\tfrac{1}{2}\psi(\tfrac{1}{2}-\tfrac{i\kappa}{4})+\gamma+2\ln 2 - 2e^{-2\omega}\frac{1+(\tfrac{\kappa}{2})^2-\cos\omega\kappa+\tfrac{\kappa}{2}\sin\omega\kappa}{1+(\tfrac{\kappa}{2})^2}\right]$$

$$\tfrac{1}{i}\sum_{n=1}^{\infty}\frac{(i\tanh\omega)^{2n-1}}{(2n-1)^{3/2}}v_{2n-1}(\kappa) \sim \frac{2}{\kappa N(\kappa)}\left[\tfrac{\pi}{2}\tanh\tfrac{\pi\kappa}{4}+e^{-2\omega}\frac{\sin\omega\kappa+\tfrac{\kappa}{2}\cos\omega\kappa}{1+(\tfrac{\kappa}{2})^2}\right].$$

The exact formulas are transcendental integrals which we will not need.

in agreement with eq.(6.14). Let's now evaluate the limits in the opposite order:

$$\lim_{\omega'\to\infty}\lim_{\omega\to\infty}\{b_\omega(\tfrac{\pi}{2}),A_{\omega'}\}_\star =$$

$$1 - \frac{1}{\pi}\lim_{\omega'\to\infty}\lim_{\omega\to\infty}\int_{-\infty}^{\infty}d\kappa\frac{e^{i\omega(\kappa-2i)}}{\sinh\frac{\pi\kappa}{2}}e^{-2\omega'}\left[1 - \frac{\cos\omega'\kappa - \frac{\kappa}{2}\sin\omega'\kappa}{1+(\frac{\kappa}{2})^2}\right]\coth\frac{\pi\kappa}{4}.$$

The integrand has a pole at $2i$ which cancels the zero of the cotangent:

$$\lim_{\omega'\to\infty}\lim_{\omega\to\infty}\{b_\omega(\tfrac{\pi}{2}),A_{\omega'}\}_\star = 1 - \frac{1}{\pi}\lim_{\omega'\to\infty}2\pi i\left[e^{-2\omega'}(-\tfrac{1}{2})(-1)\tfrac{1}{i}(\cos 2i\omega' - i\sin 2i\omega')\right]$$
$$= 1 - \lim_{\omega'\to\infty}e^{-2\omega'}(\cosh 2\omega' + \sinh 2\omega') = 0. \tag{6.15}$$

The answer is again in agreement with eq.(6.14).

## 7. The Hamiltonian

A very interesting application of the regulating procedure developed by Bars and Matsuo [15, 22, 21] is that it allows a simple representation of the Siegel gauge kinetic operator in terms of a regulated open string star product. Consider the zeroth Virasoro generator written in the mixed basis $x_{2n}\ p_{2n-1}$:

$$L_0 = \tfrac{1}{2}\sum_{n=1}^{\infty}\left[-\frac{\partial^2}{\partial x_{2n}^2} + (2n)^2 x_{2n}^2\right] + \tfrac{1}{2}\sum_{n=1}^{\infty}\left[-(2n-1)^2\frac{\partial^2}{\partial p_{2n-1}^2} + p_{2n-1}^2\right].$$

This is not normal ordered, so the action of $L_0$ on the vacuum gives a sum of the zero point energies for every string oscillator,

$$L_0\Psi_{|0\rangle} = \tfrac{1}{2}\sum_{n=1}^{\infty}n\ \Psi_{|0\rangle},$$

where,

$$\Psi_{|0\rangle} = \exp\left[-\tfrac{1}{2}\sum_{n=1}^{\infty}2nx_{2n}^2 - \tfrac{1}{2}\sum_{n=1}^{\infty}\frac{p_{2n-1}^2}{2n-1}\right].$$

The goal is to represent the action of $L_0$ algebraically using the open string star product. This is possible if the star product is regulated in such a way that $T$ is invertible. As mentioned earlier, such a regularization also seems necessary to define the discrete basis eq.(2.9). The prescription is to deform $T$ simultaneously with $v, w, R$, and the "frequency matrices" $\kappa_e, \kappa_o$ so that the relations,

$$RT = TR = 1 \quad T\bar{T} = 1 - \frac{w\bar{w}}{1+w^2} \quad R\bar{R} = 1 + (1+w^2)v\bar{v}$$
$$\bar{T}T = 1 - v\bar{v} \quad \bar{R}R = 1 + w\bar{w} \quad R = \kappa_e^{-2}T\kappa_o^2$$
$$\bar{T}w = v \quad Tv = \frac{w}{1+w^2} \quad Rw = (1+w^2)v \quad Tv = w, \tag{7.1}$$

hold with $w^2$ finite. (For simplicity we will not use new symbols to denote the regulated $T, R, w, v$.) The unregulated frequency matrices $\kappa_e, \kappa_o$ are,

$$\kappa_{o,2m-1,2n-1} = (2m-1)\delta_{2m-1,2n-1} \qquad \kappa_{e,2m,2n} = 2m\delta_{2m,2n},$$

and we assume that our regularization does not alter these expressions. One usually does not define the linear spaces on which $T$, $R$, $v$ and $w$ act/reside, since this depends in detail on how the relations eq.(7.1) are explicitly realized. Often $T$, $R$, $v$ and $w$ are chosen to be finite dimensional, though in the following we will assume that they are infinite dimensional.

A short calculation using eq.(7.1) shows that one can represent $L_0$ using the regulated star product as follows:

$$L_0 \Psi = \{\mathcal{L}_0, \Psi\}_\star + \tfrac{1}{2}(1+w^2)[[\Psi, P_L]_\star, P_L]_\star, \qquad (7.2)$$

with,

$$\mathcal{L}_0 = \tfrac{1}{4} \sum_{n=1}^{\infty} (p_{2n-1}^2 + (2n)^2 x_{2n}^2)$$

$$P_L = \tfrac{1}{2} \sum_{n=1}^{\infty} v_{2n-1} p_{2n-1}. \qquad (7.3)$$

Ideally, we would like to remove the regulator and represent $L_0$ in terms of the true star product, but the norm of $w$ diverges in this limit and the second term on the right is singular. Apparently, it is not possible to see the open string spectrum with Witten's product. What is the significance of this?

The good way to understand the role of the singular term is to see it's contribution in a concrete example. Consider the ground state wavefunctional:

$$\{\tfrac{1}{4}\sum_{n=1}^{\infty}(2n)^2 x_{2n}^2, \Psi_{|0\rangle}\}_\star = \tfrac{1}{2}\sum_{n=1}^{\infty}\left[(2n)^2 x_{2n}^2 - p_{2n-1}^2 + (2n-1)\right]\Psi_{|0\rangle}$$

$$\{\tfrac{1}{4}\sum_{n=1}^{\infty} p_{2n-1}^2, \Psi_{|0\rangle}\}_\star = \tfrac{1}{2}\left[\sum_{n=1}^{\infty}(p_{2n-1}^2 - (2n)^2 x_{2n}^2) + \frac{(\sum_{l=1}^{\infty} 2lw_{2l}x_{2l})^2}{1+w^2}\right.$$

$$\left.+ \sum_{n=1}^{\infty} 2n\left(1 - \frac{1}{1+w^2}\right)\right]\Psi_{|0\rangle}$$

$$\tfrac{1}{2}(1+w^2)[[\Psi, P_L]_\star, P_L]_\star = -\tfrac{1}{2}\frac{1}{1+w^2}\left[\left(\sum_{l=1}^{\infty} 2lw_{2l}x_{2l}\right)^2 - \sum_{n=1}^{\infty} 2n\right]\Psi_{|0\rangle}. \qquad (7.4)$$

Adding these pieces up we get the sum of string zero point energies. As $w^2 \to \infty$ the singular term makes an infinite contribution, but the infinity is "subleading" to the infinity of zero point energies.

To interpret the significance of these formulas, consider an "almost" ground state functional:

$$\Psi_\epsilon = \exp\left[-\frac{1}{\pi}\int_0^\pi d\sigma x(\sigma)\sqrt{-\frac{d^2}{d\sigma^2}}x(\sigma) + \frac{1}{\pi}\int_{\pi/2-\epsilon}^{\pi/2+\epsilon} d\sigma x(\sigma)\sqrt{-\frac{d^2}{d\sigma^2}}x(\sigma)\right],$$

where $\epsilon << 1$. The second integral cancels off the midpoint dependence of $\Psi_\epsilon$. Translating to the mixed basis gives to leading order in $\epsilon$,

$$\Psi = \exp\left[-\tfrac{1}{2}\sum_{m,n=1}^{\infty} 2n x_{2m} x_{2n}(\delta_{2m,2n} - 4\epsilon w_{2n} w_{2m}) - \tfrac{1}{2}\sum_{n=1}^{\infty} \frac{p_{2n-1}^2}{2n-1}\right].$$

Acting on this functional with $L_0$ gives,

$$L_0 \Psi = \tfrac{1}{2}\sum_{n=1}^{\infty}\left[2n + (2n-1) - 4\epsilon 2n + 4\epsilon \left(\sum_{l=1}^{\infty} 2l x_{2l} w_{2l}\right)^2\right]\Psi.$$

Identifying $\epsilon = (1+w^2)^{-1}$ we see that this is precisely the result of eq.(7.4) neglecting the contribution of the singular term. Apparently, the purpose of singular term is to resolve the zero-point energy of the string midpoint with the star product.

In ref.[17] it was observed that $L_0$ expressed in the kappa basis eq.(3.1) also appears to be somewhat singular. However, it is important to emphasize that this is a completely distinct phenomenon from the difficulty of finding a nonsingular representation of $L_0$ in terms of the star product. The "singular" expression for $L_0$ in the kappa basis comes from the fact that when it is expressed in the form $\int_0^\infty K(\kappa, \kappa') a^+(\kappa) a(\kappa)$ the kernel $K(\kappa, \kappa')$ turns out to be a complex delta function[33]. However, as explained in the last section, the complex delta function is a rigorously defined object in the topological dual of the appropriate space of analytic test functions. This space is all we need to perform precise calculations, for example to see the open string spectrum, so the apparent problems with $L_0$ in the kappa basis are really fictitious.

## 8. Conclusion

In this paper we have studied the role of the midpoint in the open string star product from several perspectives. Our hope was to illuminate some delicate issues which may ultimately be important for developing a precise understanding of the mathematical structure of string field theory, and perhaps in constructing analytic solutions.

At this point it seems appropriate to return to one of the central questions motivating our analysis: can the open string star product be interpreted as a matrix or Moyal product? The answer seems to be "almost," but no. We have seen why in some detail: the star algebra contains nontrivial states which either (anti)commute with all string fields or which vanish when multiplied with anything else. As we all know, a matrix algebra has only one element (up to a proportionality) which commutes with everything else—the identity—and only one element which gives zero when multiplied with anything else—0. Of course, since we have no precise definition of the algebra of string fields it is possible that a suitable definition would exclude commutative coordinates and null states. Barring this, the star algebra seems to be quite a bit more subtle than the matrix product intuition suggests. In retrospect this is not surprising, since the string is a single continuous, connected object, not a pair of half strings. In particular, this requires that certain local quantities on the worldsheet be either continuous or conserved through the interaction, most notably at the string midpoint, where the naive matrix product picture might allow for some sort of pathology. Remarkably, this fact alone is enough for us to have anticipated the existence of commutative and null states in the algebra: for example, since the $b$ ghost is continuous through the interaction, $b(\frac{\pi}{2})$ is identified locally should be purely (anti)commutative. On the other hand, since the momentum $p(\sigma)$ is conserved through the interaction, it must vanish at the string midpoint, implying that $p(\frac{\pi}{2})$ annihilates the vertex. So, in the end it seems we have struggled with complicated formalism to unearth a few simple facts which should in retrospect have been obvious; we nevertheless hope that the discussion has been edifying.

Some future directions may be worth exploring:

• Little work has been devoted to understanding the Moyal formulation of the open superstring star algebra and it's relation to the midpoint. One approach was developed in the continuous basis in ref.[29] but no corresponding formulation has been codified in the discrete/mixed basis.

• A clearer understanding needs to be found of issues relating to ghost/matter zero modes and the normalization of the star algebra.

• Much more work needs to be done in understanding the relationship between the Virasoro generators/BRST operator and the operator/Moyal formulation of the star product, particularly with regard to the midpoint. As yet, no one has demonstrated gauge invariance of Witten's action purely within an operator/Moyal framework. A successful demonstration would doubtless generate deep insight into the structure of string field theory.

• Finally, one can tackle the problem of defining the algebra of open string fields[18]. The basic criterion for this definition (if it exists) is that the algebra is closed under star multiplication and that Witten's axioms hold.

### Acknowledgments

I would like to thank I.Bars, D.Belov, D.Gross, A.Konechny, M. Kroyter, and D.Reynolds for many useful discussions. I especially thank M.Putinar for helping me gain a deeper appreciation of the subject from the perspective of functional analysis. This work was generously supported by the National Science Foundation under Grant No. PHY00-98395.

## A. More about the $v_n$'s

The harmonic oscillator Hamiltonian is

$$H = a^+ a + \tfrac{1}{2}.$$

---

[18]For a recent attempt to construct a definition of the star algebra, see ref.[40].

This operator has a discrete spectrum of eigenvectors $|n\rangle$ (in our conventions $1 \leq n < \infty$) corresponding to the $n - 1$st excited state of the harmonic oscillator. On the other hand, the position operator,

$$X = \frac{i}{\sqrt{2}}(a - a^+)$$

has a continuous spectrum of eigenvectors, $|x\rangle$ $(-\infty < x < \infty)$. In quantum mechanics we are often interested in the inner product,

$$\phi_n(x) = \langle x|n-1\rangle = H_n(x)\frac{e^{-x^2/2}}{(\pi 4^n (n!)^2)^{\frac{1}{4}}}.$$

The $\phi_n$'s are of course the familiar Hermite functions and the $H_n$'s are the Hermite polynomials. But in this paper we do not care so much about $X$ and its eigenvalues. More interesting is the operator[25],

$$K_1 = -\sqrt{aa^+}a^+ - a\sqrt{aa^+}.$$

$K_1$ has a continuous spectrum of eigenvectors conventionally labelled by $\kappa$ for $-\infty < \kappa < \infty$. Consider wavefunctions formed from the inner product,

$$v_n(\kappa) = \langle \kappa|n\rangle = \frac{Ч_n(\kappa)}{\sqrt{n}\sqrt{\frac{2}{\kappa}\sinh\frac{\pi\kappa}{2}}}.$$

The functions $Ч_n$ are simply a set of polynomials, like the Hermite polynomials, but orthogonal with respect to the weight function[31],

$$w(\kappa) = \left(\frac{2}{\kappa}\sinh\frac{\pi\kappa}{2}\right)^{-1}.$$

The $Ч_n$'s are unfortunately not one of the classical orthogonal polynomials we would easily find in a handbook. Therefore it's worthwhile cataloging a few of their properties:

**Differential equation**: The statement that the $v_n$'s are eigenstates of $H$ translates to the differential equation[33, 34]:

$$\sin\left(2\frac{d}{d\kappa}\right)\kappa Ч_n(\kappa) = 2n Ч_n(\kappa).$$

**Generating function[25]**:

$$\sum_{n=1}^{\infty}\frac{z^n}{n}Ч_n(\kappa) = \frac{1}{\kappa}(1 - e^{-\kappa \tan^{-1} z}) = f_\kappa(z).$$

**Recurrence relations**: The polynomials can be systematically generated from the relation

$$Ч_{n+1}(\kappa) + Ч_{n-1}(\kappa) = -\frac{\kappa}{n}Ч_n(\kappa); \quad .$$

after defining $Ч_0(\kappa) = 0$ and $Ч_1(k) = 1$. The first few polynomials are explicitly:

$$Ч_1(\kappa) = 1 \qquad Ч_4(\kappa) = -\tfrac{1}{6}\kappa^3 + \tfrac{4}{3}\kappa$$
$$Ч_2(\kappa) = -\kappa \qquad Ч_5(\kappa) = \tfrac{1}{24}\kappa^4 - \tfrac{5}{6}\kappa^2 + 1$$
$$Ч_3(\kappa) = \tfrac{1}{2}\kappa^2 - 1 \qquad Ч_6(\kappa) = -\tfrac{1}{120}\kappa^5 + \tfrac{1}{3}\kappa^3 - \tfrac{23}{15}\kappa.$$

Here is a graph of the first five $v_n(\kappa)$'s:

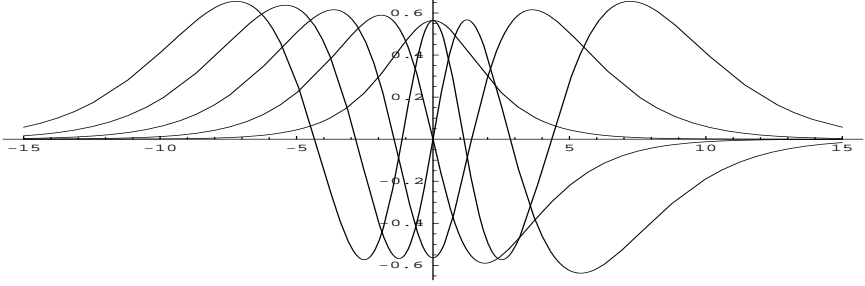

**Ladder operators**:

$$\cos^2\left(\frac{d}{d\kappa}\right)\kappa Ч_n(\kappa) = -n Ч_{n+1}(\kappa).$$

$$\sin^2\left(\frac{d}{d\kappa}\right)\kappa Ч_n(\kappa) = -n Ч_{n-1}(\kappa).$$

**Special values**:

$$Ч_{2n}(0) = 0 \qquad Ч_{2n-1}(0) = (-1)^{n+1}$$
$$Ч'_{2n}(0) = (-1)^n \sum_{m=1}^{n} \frac{1}{2m-1} \qquad Ч'_{2n-1}(0) = 0$$
$$Ч_n(2i) = n(-i)^{n-1}.$$

**Exotic Theorem:**

$$m\, i^m\, Ч_n(2im) = -n\, i^{-n}\, Ч_m(-2in).$$

**Derivative**:

$$Ч'_{2n}(\kappa) = -\frac{Ч_{2n}(\kappa)}{\kappa} + 2n \sum_{m=1}^{n} \frac{(-1)^{m-n+1}}{2n-(2m-1)} \frac{Ч_{2m-1}(\kappa)}{2m-1}$$

$$Ч'_{2n-1}(\kappa) = \frac{(-1)^{n+1} - Ч_{2n-1}(\kappa)}{\kappa} + (2n-1) \sum_{m=1}^{n-1} \frac{(-1)^{m-n}}{(2n-1)-2m} \frac{Ч_{2m}(\kappa)}{2m}.$$

**Integral representations**[33, 35]:

$$\begin{aligned}
Ч_n(\kappa) &= \frac{n}{2\pi} i^{n-1} w(\kappa) \int_{-\infty}^{\infty} du\, e^{i\kappa u} \frac{\tanh^{n-1} u}{\cosh^2 u} \\
&= \frac{n}{2\pi} i^{n-1} w(\kappa) \left( \int_1^{\infty} + \int_{-\infty}^{-1} \right) dx\, \frac{e^{i\kappa \coth^{-1} x}}{x^{n+1}} \\
&= \frac{n}{2\pi i} \oint_{z=0} dz\, \frac{f_\kappa(z)}{z^{n+1}}.
\end{aligned} \quad (A.1)$$

**Darboux-Christoffel Formula**:

$$\sum_{n=1}^{N} \frac{1}{n} Ч_n(\kappa) Ч_n(\kappa') = -\frac{Ч_{N+1}(\kappa) Ч_N(\kappa') - Ч_{N+1}(\kappa') Ч_N(\kappa)}{\kappa - \kappa'}.$$